\documentclass[nocopyrightspace]{sigplanconf}

\usepackage{amsmath}
\usepackage{listings}
\usepackage{graphicx}
\usepackage{dcolumn}
\usepackage{bm}

\usepackage{alltt}
\usepackage{moreverb}
\usepackage{tikz}
\usetikzlibrary{positioning,shapes,shadows,arrows}

\usepackage{color}

\newcommand{\Liquid}{LIQ$Ui\ket{}$\ }
\newcommand{\LiquidT}{LIQ$\bm{Ui\ket{}}$}
\newcommand{\LiquidB}{LIQ$Ui\ket{}$}

\usetikzlibrary{
arrows,shapes.misc,shapes.arrows,matrix,
positioning,scopes,decorations.pathmorphing,shadows,
shapes.geometric,decorations.pathreplacing,shapes.gates.logic.US,calc}

\tikzset{
matStyle/.style={column sep=1mm,row sep=1mm,execute at empty cell=\node{\phantom{.}};},
measStyle/.style={and gate US,text centered,minimum size=6mm,thick,draw=black,
    top color=white,bottom color=green!10!white,font=\itshape},
meterStyle/.style={rectangle,text centered,minimum size=6mm,thick,draw=black,
    top color=white,bottom color=green!10!white,font=\itshape},
lblStyle/.style={rectangle,minimum size=6mm,thick,draw=white,
	top color=white,bottom color=white,font=\itshape},
prepStyle/.style={rectangle,minimum size=6mm,thick,draw=white,text=red!60!black,
	top color=white,bottom color=white,font=\itshape},
ctrlStyle/.style={circle,text centered,inner sep=-1pt,minimum size=1mm,thick,draw=black,
	top color=black,bottom color=black},
ctrloStyle/.style={circle,text centered,inner sep=-1pt,minimum size=1mm,thick,draw=black,
	top color=white,bottom color=white},
targStyle/.style={circle,text centered,inner sep=0pt,minimum size=3mm,thick,draw=black,
	top color=white,bottom color=white,append after command={
        [shorten >=\pgflinewidth, shorten <=\pgflinewidth, thick,]
        (\tikzlastnode.north) edge[thick] (\tikzlastnode.south)
        (\tikzlastnode.east) edge[thick] (\tikzlastnode.west)
        }},
gateStyle/.style={rectangle,minimum size=6mm,thick,draw=black,
	top color=white,bottom color=blue!10!white},
mgStyle/.style={rectangle,minimum size=6mm,transparent},
mgBoxStyle/.style={rectangle,minimum size=6mm,thick,draw=black,
	top color=white,bottom color=blue!10!white},
qwStyle/.style={thick,draw=black,font=\ttfamily},
qwxStyle/.style={thick,draw=black,font=\ttfamily},
>=latex,thick,
/pgf/every decoration/.style={/tikz/sharp corners},
pointStyle/.style={coordinate},
>=stealth',thick,draw=black,tip/.style={->,shorten >=0.007pt},
every join/.style={rounded corners},
hv path/.style={to path={-| (\tikztotarget)}},
vh path/.style={to path={|- (\tikztotarget)}},
text height=1.5ex,text depth=.25ex
}

\pgfdeclarelayer{background}		
\pgfdeclarelayer{foreground}		
\pgfsetlayers{background,main,foreground}

\newcommand{\gate}[2][] {\node (#1) [gateStyle]{$#2$};}
\newcommand{\lbl}[2][]{\node (#1) [lblStyle] {$#2$};}

\newcommand{\mg}[2][]{\node (#1) [mgStyle] {\phantom{$#2$}};}
\newcommand{\point}[1][]{ \node (#1) [pointStyle] {};}
\newcommand{\targ}[1][]{ \node (#1) [targStyle] {};}
\newcommand{\ctrl}[1][]{ \node (#1) [ctrlStyle] {};}

\newcommand{\meter}[1][meter]{
    \node (#1) [meterStyle] {
    	\begin{tikzpicture}
    	\phantom{.}
    	  \draw +(0,-.14) arc (0:180:.2 and .16);
    	  \draw[thick] (-0.2,-.14) -- (-0.1,+.13);
	\end{tikzpicture}
    };
}
\newcommand{\multigate}[4][]{
    \draw[gateStyle,name=#1] (#2.north west) rectangle (#3.south east);
    \draw[draw=black] ($0.5*(#2)+0.5*(#3)$) node{$#4$};
}

\newcommand{\qw}[2]{\draw[qwStyle] (#1) -- (#2);}
\newcommand{\cw}[2]{\draw[qwStyle, double] (#1) -- (#2);}

\iftrue 
    \newcommand{\qwx}[2]{\draw[qwxStyle] (#1) -- (#2);}
    \newcommand{\cwx}[2]{\draw[qwxStyle, double] (#1) -| (#2);}
    \newcommand{\dwx}[2]{\draw[qwxStyle, dotted] (#1) -- (#2);}
\else 
    \newcommand{\qwx}[2]{
        \coordinate (rnd) at (canvas cs:x=3pt*rand);
        \draw[qwxStyle] ($(#1)+(rnd)$) -- ($(#2)+(rnd)$);
    }
    \newcommand{\cwx}[2]{
        \coordinate (rnd) at (canvas cs:x=3pt*rand);
        \draw[qwxStyle, double] ($(#1)+(rnd)$) -| ($(#2)+(rnd)$);
    }
    \newcommand{\dwx}[2]{
        \coordinate (rnd) at (canvas cs:x=3pt*rand);
        \draw[qwxStyle, dotted] ($(#1)+(rnd)$) -| ($(#2)+(rnd)$);
    }
\fi

\newcommand{\ket}[1]{{\left\vert{#1}\right\rangle}}

\begin{document}

\tikzstyle{box}=[rectangle,dotted,draw=black, rounded corners, drop shadow,text centered, anchor=north, text=black, text width=2cm,scale=0.8]
\tikzstyle{styMat}=[fill=gray!40,matrix anchor=north]
\tikzstyle{styGrn}=[box,fill=green!80]
\tikzstyle{styYel}=[box,fill=yellow!80]
\tikzstyle{styAqu}=[box,fill=blue!20!yellow]
\tikzstyle{styBlu}=[box,fill=blue!40]
\tikzstyle{styRed}=[box,fill=red!40]
\tikzstyle{styOra}=[box,fill=orange!40]
\tikzstyle{myarrow}=[->,>=stealth,very  thick]

\setlength{\pdfpageheight}{\paperheight}
\setlength{\pdfpagewidth}{\paperwidth}


\conferenceinfo{CONF 'yy}{Month d--d, 20yy, City, ST, Country}
\copyrightyear{20yy}
\copyrightdata{978-1-nnnn-nnnn-n/yy/mm}
\doi{nnnnnnn.nnnnnnn}




\titlebanner{banner above paper title}        
\preprintfooter{short description of paper}   

\title{\LiquidT: A Software Design Architecture and Domain-Specific Language for Quantum Computing}
\subtitle{}

\authorinfo{Dave Wecker $^*$\and Krysta M. Svore\titlenote{Quantum Architectures and Computation Group, Microsoft Research, One Microsoft Way, Redmond, WA, 98052}}
           {Microsoft Research}
           {\{wecker,ksvore\}@microsoft.com}

\maketitle

\begin{abstract}
Languages, compilers, and computer-aided design tools will be essential for scalable quantum computing, which promises an exponential leap in our ability to execute complex tasks.
\Liquid is a modular software architecture designed to control quantum hardware.
It enables easy programming, compilation, and simulation of quantum algorithms and circuits, and is independent of a specific quantum architecture.
\Liquid contains an embedded, domain-specific language designed for programming quantum algorithms, with F\# as the host language.
It also allows the extraction of a circuit data structure that can be used for optimization, rendering, or translation.
The circuit can also be exported to external hardware and software environments.
Two different simulation environments are available to the user which allow a trade-off between number of qubits and class of operations.
\Liquid has been implemented on a wide range of runtimes as back-ends with a single user front-end.
We describe the significant components of the design architecture and how to express any given quantum algorithm.
\end{abstract}







\section{Introduction}\label{sec:Intro}

The harnessing of quantum mechanics for computation will cause a paradigm shift in our notions of computational methods and devices.
In recent years, we have seen problems in mathematics and computer science for which a quantum algorithm is exponentially faster than the best-known classical algorithm.
Problems include factoring integers \cite{Shor1994}, estimating the ground state energy of complex molecules \cite{AG2005,Chem2009}, and solving systems of linear equations \cite{HHL2009}.
The pursuit of harnessing the laws of quantum physics for computational speed-ups is both challenging and rewarding.
Problems solvable more quickly on a quantum computer are only beginning to be unveiled and the need for a high-level programming environment to aid in their development is apparent.

There is a long history of creating software languages that encourage higher-level abstractions freeing the user to focus more on problem solving and less on the details of the specific hardware involved.
Probably the best example of this was the introduction of FORTRAN \cite{Fortran1957} that moved an entire community away from machine code and into general purpose algorithm creation.
Quantum computing is no exception.
Quantum computing possesses unique computational attributes which require novel programming constructs to enable harnessing and manipulation of quantum states.
One of the grand challenges for the computer science and programming language community will be the design and implementation of a system architecture to control quantum hardware. \Liquid is an evolutionary step along the way, building on previous language formalisms.

A software design architecture for quantum computing \cite{Svore2006} should offer high-level abstractions of quantum physics and linear algebra as well as the automation of complex tasks for easy development, simulation, and testing of quantum algorithms.
The quantum programming language needs to allow a description of any quantum circuit at a suitable level of abstraction.
It must include both quantum primitives as well as classical control definitions.
Finally, it should offer an environment which aids in the understanding of quantum physics, provides easy manipulation of quantum circuits and classical control, and allows development of large-scale quantum algorithms for ultimate deployment on a quantum computer.

Current state-of-the-art software architectures for quantum computing lack tools for control of quantum hardware and scalable quantum algorithm development.
Most research is focused on developing circuits for small subroutines of quantum algorithms and performing resource cost estimates.
In contrast, \Liquid (which stands for ``Language Integrated Quantum Operations"\footnote{A quantum operation is usually referred to as a unitary operator ($U$) applied to a column state vector (also known as a ket: $\ket{\cdot}$). The “$i$” is just a constant scaling factor, hence the acronym.}) is an attempt to provide users with an end-to-end exploration and control environment from algorithm writing, to visualization, to simulation, emulation, and deployment on target hardware.
%

The ultimate goal behind \Liquid is to control quantum hardware.
\Liquid contains a robust, large-scale domain-specific language embedded in F\# and isolated runtime for programming quantum algorithms.  It contains modular tools for circuit manipulation, simulation, export, and rendering.
In addition, it has the ability to support investigations of quantum noise, quantum error-correcting codes (QECC), circuit decomposition and optimization, classical control integration, and architecture-specific timing and layout constraints.


We organize our presentation of \Liquid as follows.
In Section \ref{sec:Related}, we review several existing quantum programming languages and their similarities and differences to \LiquidB.
In Section \ref{sec:QC}, we provide a brief background on the primitives of quantum computation and quantum algorithm design.
We introduce the \Liquid software design architecture in Section \ref{sec:liquid} and describe the primary elements of our system, including the language, simulators, and backends.
We provide several code examples in Section \ref{sec:Prog}.
In Section \ref{sec:Shor}, we show how to program and simulate Shor's algorithm in \LiquidB.
Finally, we conclude and discuss future directions in Section \ref{sec:Conclude}.

\section{Related Work}\label{sec:Related}

Several quantum programming languages have been proposed in recent years~\cite{Miszczak2011}.
%
%
Quantum Computation Language (QCL) \cite{Omer1998,Omer2000,Omer2003} is perhaps the most advanced imperative quantum programming language.
It is a C-style language designed for easy, structured programming and natural quantum algorithm design.
QCL divides the components of a quantum algorithm into ``quantum functions" (unitary operations), ``pseudo-classical operators" (quantum oracles), and ``classical procedures" (classical operations).

Another imperative quantum programming language called Q Language was proposed by Betelli et al.~\cite{Bettelli2003}.
It allows simulation of decoherence on a quantum algorithm, which is especially important since quantum computers are inherently noisy and in their infancy.
Q is developed as a class library for C++ and provides classes for basic quantum gates.  
The class of gates is also user-extensible.
Both QCL and Q lack quantum data types and formal semantics.

Functional languages for quantum programming have also been proposed.
The quantum lambda calculus was originally developed in the form of a simulation library for the Scheme language \cite{Tonder2004} and later became an ML-style language with strong static type checking \cite{Selinger2006,Selinger2009}.
Although rigorous, the quantum lambda calculus lacks facilities for construction and manipulation of quantum circuits.
%
The Quantum IO Monad \cite{Altenkirch2010} is embedded in Haskell and offers consistent operational semantics.
However, it lacks suitable design tools for development of quantum algorithms.
\Liquid is a reduction of ideas drawn from these formal functional languages to a practical, user-friendly system that enables the development of quantum algorithms and the programming of quantum devices.

Recently, Quipper has been introduced as a language to enable high-level programming of scalable quantum computations \cite{Selinger2013}.
Quipper is a strongly-typed, functional quantum programming language embedded in Haskell.
Both Quipper and \Liquid offer powerful and extensible facilities for quantum circuit description and
manipulation, including gate decomposition and circuit optimization; both include classical components such as measurements and classically-controlled gates; both offer a way to represent algorithms and circuits at multiple levels of abstraction; both systems allow quantum circuits to be exported for rendering or resource costing; and both systems are modular and user-extensible. 
However, the exact implementation details between the two systems differ.

In contrast to Quipper, we have designed \Liquid explicitly with quantum hardware in mind. 
We believe that the model of quantum computation closely matches the traditional model of a co-processor.
Qubits are real entities that have lifetimes and are mutable. 
In \LiquidB, the qubit type reflects this reality.
\Liquid also does not have built-in gates. 
All gates are implemented within a library which can be modified or replaced by the user. 

While both systems are equipped with simulators for universal quantum circuits, as well as more efficient specialized simulators for stabilizer and other classes of circuits, \LiquidB's simulators are highly optimized, taking advantage of many available techniques, including custom memory management, cache coherence analysis, parallelization, ``gate growing", and virtualization (running in the cloud).
\LiquidB's highly optimized simulation environment allows thorough investigation of quantum algorithms under noise, physical device constraints, and simulation.

\Liquid is also a full optimizing compiler. 
A user's input circuit definition may be massively rewritten (under user control) to generate compact, highly-optimized versions for simulation.
We can compile any given unitary circuit with varying levels of optimization and can mathematically prove that the pre- and post-optimized unitary are identical even though the resulting circuits may appear very different.
Another unique component of \Liquid is its ability to perform Hamiltonian
simulations, including the efficient simulation of Trotterized
circuits, as well as computations in the adiabatic model of quantum
computation.

\section{Quantum Computation}\label{sec:QC}
In this section, we briefly review primitives of quantum computation.
A detailed review can be found in \cite{Nielsen2000}.

\subsection{Qubits and Quantum Gates}
In quantum computation, quantum information is stored in a quantum bit, or \textit{qubit}.
Whereas a classical bit has a state value $s\in \{0,1\}$, a qubit state $\ket{\psi}$ is a \textit{linear superposition} of states:
\begin{equation}
\ket{\psi} = \alpha\ket{0} + \beta\ket{1} = \left[
    \begin{smallmatrix}
    \alpha \\
    \beta \end{smallmatrix} \right] \label{eqn:qubit},
\end{equation}
where the $\{0,1\}$ basis state vectors are represented in Dirac notation (called \textit{ket} vectors) as
$\ket{0} = \begin{bmatrix}\setlength{\arraycolsep}{3pt} 1 & 0 \end{bmatrix}^T$, and
$\ket{1} = \begin{bmatrix} 0 & 1 \end{bmatrix}^T$, respectively.
The {\it amplitudes} $\alpha$ and $\beta$ are complex numbers that satisfy the normalization condition: $|\alpha|^2 + |\beta|^2 = 1$.
Upon {\it measurement} of the quantum state $\ket{\psi}$, either state $\ket{0}$ or $\ket{1}$ is observed with probability $|\alpha|^2$ or $|\beta|^2$, respectively.

An $n$-qubit quantum state lives in a $2^n$-dimensional Hilbert space and is represented by a $2^n \times 1$-dimensional state vector whose entries represent the amplitudes of the basis states.
A superposition over $2^n$ states is given by:
\begin{equation}
\ket{\psi} = \sum_{i=0}^{2^n - 1} \alpha_i \ket{i}\mbox{, such that }\sum_i |\alpha_i|^2 = 1 \label{eqn:ket},
\end{equation}
where $\alpha_i$ are complex amplitudes and $i$ is the binary representation of integer $i$.
Note, for example, that the three-qubit state $\ket{000}$ is equivalent to writing the tensor product of the three states:
$\ket{0}\otimes\ket{0}\otimes\ket{0} = \ket{0}^{\otimes 3} = \left[\begin{smallmatrix} 1 & 0 & 0 & 0 & 0 & 0 & 0 & 0 \end{smallmatrix}\right]^T$.
The ability to represent a superposition over exponentially many states with only a linear number of qubits is one of the essential ingredients of a quantum algorithm --- an innate massive parallelism.

In a quantum computation, a closed quantum system transforms by \textit{unitary} evolution.
In particular, the quantum state $\ket{\psi_1}$ of the system at time $t_1$ is related to the quantum state $\ket{\psi_2}$ at time $t_2$ by a unitary operator $U$ that depends only on $t_1$ and $t_2$:
\begin{equation}
\ket{\psi_2} = U\ket{\psi_1}
\label{eqn:op}
\end{equation}
In turn, quantum operations are necessarily \textit{reversible}.
We refer to quantum unitary operations as quantum \textit{gates}.
Measurement is not reversible;
it collapses the quantum state to the observed value, thereby erasing the knowledge of the amplitudes $\alpha$ and $\beta$.

An $n$-qubit quantum gate is a $2^n \times 2^n$ unitary matrix that acts on an $n$-qubit quantum state.
For example, the {\it Hadamard} gate {\tt H} maps
$\ket{0} \rightarrow \frac{1}{\sqrt{2}}\left( \ket{0} + \ket{1}\right)$, and
$\ket{1} \rightarrow \frac{1}{\sqrt{2}}\left( \ket{0} - \ket{1}\right)$.
The {\tt X} gate, similar to a classical NOT gate, maps
$\ket{0}\rightarrow \ket{1}$, and
$\ket{1}\rightarrow \ket{0}$.
The {\tt Z} gate maps $\ket{1}\rightarrow -\ket{1}$.
The identity gate is represented by {\tt I}.
The two-qubit {\it controlled-}NOT gate, {\tt CNOT}, maps $\ket{x,y}\rightarrow\ket{x, x\oplus y}$.
The corresponding unitary matrices are:
 \[
\mbox{H} =
  \left[\begin{smallmatrix}1&1\\1&\textrm{-}1\end{smallmatrix}\right],
  ~ \mbox{\tt X} =
  \left[\begin{smallmatrix}0&1\\1&0\end{smallmatrix}\right],
~ \mbox{\tt Z}  =
  \left[\begin{smallmatrix}1&0\\0&\textrm{-}1\end{smallmatrix}\right],
  ~ \mbox{\tt I} = \left[\begin{smallmatrix}1&0\\0&1\end{smallmatrix}\right],
  \mbox{\tt CNOT} = \left[\begin{smallmatrix}1 & 0 & 0 & 0 \\ 0 & 1 & 0 & 0 \\ 0 & 0 & 0 & 1 \\ 0 & 0 & 1 & 0\end{smallmatrix}\right].
\]

Quantum state evolution is represented in a \textit{quantum circuit diagram}, where time flows from left to right.
Solid wires represent qubits; double wires represent classical bits.
Single-qubit gates are represented by boxes containing their symbol.
{\tt CNOT} is denoted by a vertical line between a $\bullet$ (to represent the control qubit in state $\ket{1}$) and a $\oplus$ (to represent XOR).
Measurement is denoted by the meter symbol.

\subsection{Example}
A remarkable example of quantum computation is quantum teleportation \cite{Nielsen2000}.
It enables moving quantum information around without access to a quantum communications channel.
The goal is for a messenger $M$ to deliver a source qubit (\textit{src}) to a recipient $R$ with perfect fidelity using very little classical communication.
$M$ does not know the value of the source qubit and is only allowed to send \textit{classical} information to $R$.
Quantum teleportation highlights several important primitives of quantum computation, including superposition, entanglement, and classically-controlled quantum gates.

\begin{figure}[ht]
\centering
\begin{tikzpicture}[scale=1.00,every node/.style={scale=0.80}]
\matrix[matStyle] {
  \lbl[0-0]{Src} & \point[0-1] &  &  & \ctrl[0-4] & \gate[0-5]{H} & \meter[0-6] &  & \ctrl[0-8] &  & \point[0-10]  \\
  \lbl[1-0]{\ket{0}} & \point[1-1] & \gate[1-2]{H} & \ctrl[1-3] & \targ[1-4] &  & \meter[1-6] & \ctrl[1-7] &  &  & \point[1-10]  \\
  \lbl[2-0]{\ket{0}} & \point[2-1] &  & \targ[2-3] &  &  &  & \gate[2-7]{X} & \gate[2-8]{Z} & \lbl[2-9]{Dest} & \point[2-10]  \\
   &  &  &  &  &  &  &  &  &  &   \\
};
\begin{pgfonlayer}{background}
\qw{0-1}{0-6} \qw{1-1}{1-6} \qw{2-1}{2-10} \cw{0-6}{0-10} \cw{1-6}{1-10} \qwx{0-4}{1-4}
\qwx{1-3}{2-3} \cwx{0-8}{2-8} \cwx{1-7}{2-7}
\end{pgfonlayer}
\end{tikzpicture}
\caption{\label{fig:TeleDraw}Teleportation circuit auto-generated by \LiquidB.}
\end{figure}
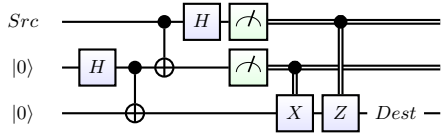

The quantum circuit for teleportation is shown in Figure \ref{fig:TeleDraw}.
The protocol begins with messenger $M$ and recipient $R$ each having a qubit in state $\ket{0}$ (bottom two qubits).
They entangle their qubits to create an EPR pair: $M$ applies a {\tt H} gate followed by a {\tt CNOT} between the two qubits.
$M$ and $R$ then travel arbitrarily far apart from each other, taking their respective qubit with them.
$M$ is then given a message qubit (\textit{src}) to send to $R$.  She entangles \textit{src} with her half of the EPR pair using a {\tt CNOT} and {\tt H} gate.
$M$ then measures \textit{src} and her half of the EPR pair and sends two (classical) measurement results to $R$ over a classical channel.
$R$ looks at the two values and conditionally applies an {\tt X} and/or {\tt Z} gate to his half of the EPR pair (bottom qubit).
The state of his qubit, labeled \textit{dest}, is now equal to the original \textit{src} state.
When a variant of this circuit is run in reverse, it can be used to perform \textit{quantum superdense coding} \cite{Nielsen2000}.

\subsection{Model of Computation}
Unlike classical computation, quantum computation is inherently \textit{probabilistic} due to measurement.
To read the output of a quantum algorithm or circuit as a classical bit string, the final quantum state is measured, which probabilistically projects the state onto one of the computational basis states.
The interplay between the quantum circuit and classical control necessitates a \textit{hybrid architecture} employing both quantum hardware and a classical computer.
Feedback between classical and quantum hardware is required for, e.g., classical control instructions, conditional circuit application, measurement, and classical pre- and post-processing of the input and output of the quantum device.

Several formal models of quantum computation have been proposed, including the quantum Turing machine \cite{Deutsch1985}, the quantum circuit model \cite{Yao1993}, the quantum adiabatic model \cite{Farhi}, and the quantum random access machine (QRAM) \cite{Knill1996}.
The quantum circuit model \cite{Yao1993} allows the representation of actual physical operations performed in the laboratory as a circuit, but does not provide definitions for classical control instructions which are required to express a given quantum algorithm.
The QRAM model extends the circuit model to include definitions for universal quantum and classical computation, including classically-controlled quantum operations.  It also includes the notion of quantum registers containing qubits.

\Liquid is developed around the quantum circuit and QRAM models, employing the quantum circuit as its underlying representation.
We adhere to the quantum circuit representation since it is universal and allows us to emulate other quantum models of computation easily. For example, we have defined interfaces that allow the user to do adiabatic evolution on first-quantized Hamiltonians and Trotter simulation on second quantized Hamiltonians.
\Liquid assumes hardware independence; it does not rely on a specific classical or quantum hardware architecture.
It assumes quantum operations can be performed in parallel if they act on distinct sets of qubits, where the amount of parallelism is subject to the constraints of the targeted quantum device.
It also allows a sequence of quantum gates to be conditionally applied based on the output of earlier quantum measurements.

\Liquid allows export to target-specific devices and simulators.
Its circuit manipulation modules are user extensible to allow translation to hardware-specific gate instructions.
Since real control of a quantum computer will be highly susceptible to noise, \Liquid enables investigation of quantum noise models at the circuit and device levels.
An example is given in Section \ref{sec:QECC}.

\subsection{Quantum Algorithm Design}

Many quantum algorithms have been proposed in recent years \cite{JordanZoo}.
Typically, they are described at the level of mathematics and physics, as opposed to at the level of quantum circuits.
However, the algorithm can be mapped to a quantum circuit, resulting in components such as state preparation, classical pre- and post-processing, quantum subroutines, quantum oracles, and measurement.
Some quantum algorithms are more easily expressed in the quantum adiabatic model \cite{Farhi}, which may also be implemented in \Liquid using a first-quantized Hamiltonian representation.

At the beginning of a quantum algorithm, \textit{quantum state preparation} is performed to initialize the quantum states.  
States are initialized to $\ket{0}$ and a quantum circuit can be applied to transform the value of the quantum register.
Qubits used as ``scratch space" are called ancilla qubits and initialized to $\ket{0}$ states.
Ancilla qubits can be reset to $\ket{0}$ during the computation to allow later reuse.

A quantum algorithm typically uses of one or more common quantum subroutines, such as
\textit{amplitude amplification} for increasing the amplitude of a desired state in a quantum superposition,
\textit{quantum phase estimation} for estimating eigenvalues of a unitary operator, and
the \textit{quantum Fourier transform} for performing a change of basis analogous to the classical discrete Fourier transform.
Manipulations of the quantum subroutine may include the \textit{reversal} or \textit{adjoint} to ``undo" a computation or reset an ancilla qubit to $\ket{0}$, and \textit{repetition} to increase precision (e.g., amplitude amplification).
All such subroutines and manipulations can be expressed as quantum circuits and are available in \Liquid (see Section \ref{sec:Functions}).

Many quantum algorithms rely on a \textit{quantum oracle} to perform function evaluation.
Recall that a classical oracle is normally a boolean function that maps an $n$-dimensional boolean input to an $m$-dimensional boolean output.
An algorithm then queries the oracle to perform the mapping from $n\rightarrow m$.
A classical boolean oracle can be converted into a quantum oracle by increasing the input and output spaces from $n$ and $m$ bits, respectively, to $n+m$ qubits each.
This maps the boolean function to a reversible function that can be represented as a unitary matrix.
Example oracles include arithmetic functions, graph functions, and lookup tables.
\Liquid supports the definition of quantum oracles; an example is given in Section \ref{sec:Shor}.

To read the output of the algorithm as a classical bit string, the final quantum state is \textit{measured} and optionally post-processed classically.
\textit{Classical pre- and post-processing} consists of classical manipulation of the data before input or after measurement.
Classical pre-processing is a series of classical procedures performed prior to initialization of the quantum states.
Classical post-processing includes checking if the classical output is a proper solution to the problem (when a solution can be efficiently verified), performing statistical analysis of output,
and determining when the algorithm can be terminated.
Classical processing can also be heavily interleaved with the logical operations of the quantum algorithm.
For example, during quantum error correction, the next sequence of quantum operations is determined based on the error syndrome measurement outcomes.

\section{\LiquidT Software Design Architecture}\label{sec:liquid}

A software architecture for scalable quantum computing requires programming languages, compilers, optimizers, and simulators with well-defined interfaces between the components \cite{Svore2006}.
We have architected \Liquid with the components of a desirable quantum design architecture in mind.
The various system components have been carefully designed to promote efficiency and interoperability.
Since the input and output formats are modular, interoperability with other tools, languages, or operating systems is easily achievable.
For example, we can currently import data from classical quantum chemistry systems (orbital integrals) and can output state vectors, circuits, and compiled unitaries for export to large linear algebra packages if desired.
\begin{figure}[t]
\centering
\begin{tikzpicture}[node distance=0.1cm]]

\matrix(matLang)[styMat,every node/.style={styGrn}] {
\node[very thick,solid] {\textbf{Language:}}; & \node {F\#}; \\
\node {Script}; & \node{C\#}; & & \\
} ;
\matrix(matGate)[styMat,below =0.2 of matLang,every node/.style={styYel}] {
\node[very thick,solid] {\textbf{Functions:}}; & \node {Gates...}; & \\
} ;

\matrix(matCirc)[styMat,below =0.2 of matGate,every node/.style={styRed}] {
\node[very thick,solid] {\textbf{Circuits:}}; & \node {Optimize}; & \node {QECC}; \\
\node {Repl}; & \node{Export}; & \node{Render...}; \\
} ;

\matrix(matSim)[styMat, below = 0.2 of matCirc,every node/.style={styAqu}] {
\node[very thick,solid] {\textbf{Simulators:}}; & \node {Universal}; \\
\node {Stabilizer}; & \node {Hamiltonian}; \\
} ;

\matrix(matRun)[styMat, below =0.2 of matSim,every node/.style={styBlu}] {
\node[very thick,solid] {\textbf{Runtimes:}}; & \node {Client}; \\
\node {Service}; & \node {Cloud}; \\
} ;

\matrix(matExp)[styMat, below =0.2 of matRun,every node/.style={styOra}] {
\node[very thick,solid] {\textbf{Backends:}}; & \node {Classical}; \\
& \node {Quantum}; \\
} ;

\draw[myarrow] (matLang.south) -- (matGate.north);
\draw[myarrow] (matGate.south) -- (matCirc.north);
\draw[myarrow] (matCirc.south) -- (matSim.north);
\draw[myarrow] (matSim.south) -- (matRun.north);
\draw[myarrow] (matGate.west) -- ++(-1.0,0) |- (matSim.west);
\draw[myarrow] (matCirc.east) -- ++(0.5,0) |- (matExp.east);

\end{tikzpicture}
\caption{\label{fig:arch}\Liquid architecture.}
\end{figure}

The \Liquid software architecture is summarized in Fig.~\ref{fig:arch}.
Entry into the system is via a programming language.
This can be F\# via a compilation environment, the F\# interpreter, or any other high-level language (e.g., C\#) that has the ability to link with the \Liquid library.
This input is then either compiled to run directly on a simulator, or is sent to a circuit manipulator that edits the circuit in a desired way and prepares it for either simulation, export, or drawing.
The available simulators run on Windows Clients and Servers, as a Windows Service on any machines that share a LAN or Cluster, and in the Azure cloud for remote execution.

\subsection{Language}\label{sec:Lang}

A quantum programming language should preserve the ``no-cloning" theorem \cite{Nielsen2000} which says that an arbitrary quantum state cannot be copied.
It must also support the unitary evolution of a quantum state.
In some languages, a qubit is viewed as a linear type (e.g., to preserve no cloning) and the development of a full logic based on this approach is attempted in \cite{Selinger2006,Selinger2009}.
However, a linear type does not map well to the reality of a physical qubit which is a truly mutable entity and may be read classically in specific circumstances without destroying it (non-destructive measurement).
A language based on linear types is also challenging to both implement and write programs in.
A functional language with an isolated physical model (further described below), on the other hand, offers efficient, compact code and static type-checking.

We have chosen to develop a domain-specific language embedded in F\#.
We chose F\# because it provides .NET support including classes and supports both object-oriented and functional programming.  It allows introspection of compiled code (reflection), access to the compiler's internal Abstract Syntax Tree (AST), and strong static typing.
F\# also provides an easy interface to external libraries (including non-managed C++ and FORTRAN math libraries) and can be called by other languages.
It supports the full suite of Microsoft Visual Studio development tools including multi-threaded debugging and performance analysis.

\Liquid relies on a basic set of data types that are embedded in the host language. These include:
\begin{description}
\itemsep 1pt
\parskip 0pt
  \item[\tt Bit] A classical value that may take values $\{{\tt Zero}, {\tt One}, {\tt Unknown}\}$.  {\tt Unknown} represents the value of a {\tt Qubit} that has not been measured (still in a quantum state).
  \item[\tt Qubit] A quantum value defined by Eq~(\ref{eqn:qubit}). Its {\tt Bit} value moves from {\tt Unknown} to {\tt Zero} or {\tt One} after measurement.
  \item[\tt Ket] The state vector representing all qubits in our system, as defined in Eq~(\ref{eqn:ket}).
This starts at the size of the number of qubits $n$ in the system, but as qubits become entangled, it may grow as large as $2^n$. Efficient handling of the state vector and operations on it is one of the central roles of \LiquidB.
  \item[\tt Gate] To perform operations on {\tt Ket}s, we define {\tt Gate}s.
In its simplest form it may be a unitary matrix that defines a specific operation (e.g., {\tt H}, {\tt X}, {\tt CNOT}, \ldots) as defined in Eq~(\ref{eqn:op}).
There are also non-unitary {\tt Gate}s (e.g., Measurement and Reanimation) as well as several meta operations.
  \item[\tt Operation] Since {\tt Gate}s are merely data structures, when wrapped in F\# functions they become operators that can apply a {\tt Gate} to a set of {\tt Qubit}s.
Calling the Hadamard function ({\tt H}) on a list of qubits ({\tt qs}) in F\# merely becomes ``{\tt H qs}'', where {\tt H} is applied to the first qubit in the list.
A {\tt Gate} will generate an error if handed arguments that do not match its definition.
  \item[\tt Circuit] One of the goals of \Liquid is to provide various manipulations of quantum algorithms such as drawing, parallelizing, substitution (some gates will not be available in target physical systems), optimization, export, and re-execution.
The {\tt Circuit} data structure achieves this goal.
Instead of running the {\tt Operation}s defining the quantum algorithm, the same calls can be used to build a {\tt Circuit} that can be manipulated by a variety of tools.
\end{description}

By design, {\tt Qubit}s and {\tt Ket}s are implemented as \textit{opaque} types that can only interact with the rest of the system via defined interfaces (creation, properties, and functions) that are restricted to operate within the scope of the opaque type (a monad).
It has no access to program state outside of itself and is opaque to the user.
This allows these types to be state-full and at the same time not pollute the rest of the functional environment.
{\tt Qubit}s and {\tt Ket}s are objects that exist on their own and may be communicated with, but are fully isolated from the rest of the system.

In addition, {\tt Qubit}s are merely identifiers to refer to parts of the state vector ({\tt Ket}). The {\tt Ket} contains all information about the simulation and {\tt Gate}s are applied to the {\tt Ket} one at a time (via direct function calls) or as an extracted {\tt Circuit} data structure (reflected on from the function calls that would have been done).
This data structure may be manipulated in many ways, as described in the next two sections, but is ultimately viewed as a sequence of {\tt Gates} applied to the {\tt Ket}.

There is not a static global ordering of the {\tt Qubit}s in a {\tt Ket} state vector. 
The arguments to the gate definitions are user-defined and are local to that function's definition. 
{\tt Qubit}s themselves are self-identifying and unique. 
If higher-level abstractions are desired, they can be easily user-defined (e.g., quantum registers).

The use of lists in {\tt Gate} definitions is a choice. 
{\tt Gate}s can equally be defined using strongly-typed, fixed or variable arguments.
In fact, all arguments are required to be strongly typed in the system.
{\tt Qubit}s, {\tt Ket}s, {\tt Gate}s, and {\tt Circuit}s are all strong quantum types and enforced within the system.
\Liquid gates are functions. The function can be as high- (or low-) level as the user desires.

\Liquid has the capability to create and destroy {\tt Qubit}s, for example for ancilla allocation, however we do not currently provide a programmatic interface to the user. We are exposing this in a future version. Note that the classical programming language is unrestricted and that {\tt Gate}s may contain any number of local classical variables.

\subsection{Functions}\label{sec:Functions}

Executable {\tt Gate}s used in a quantum algorithm are referred to in \Liquid as
{\tt Operation}s. An {\tt Operation} appears externally as a typical F\# function whose
signature is required to have the last argument as a list of qubits and
returns {\tt unit} (void). The qubits are required to define where the gate
operates within a state vector ({\tt Ket}) and since {\tt Qubit}s/{\tt Ket}s exist in
their own scope, the function never returns a value. An {\tt Operation} can be unitary or non-unitary.

One of the unique aspects of \Liquid is that all {\tt Operation}s/{\tt Gate}s are user
functions/class instances which may be extended by the user as
desired. To allow this, the {\tt Gate} class also defines instructions for how it
is to be rendered and run-time aspects that are needed by the system.
This makes the system fully extensible.

Measurement, written as {\tt M}, represents a non-unitary gate.
It is a special case of a {\tt Gate} that causes the collapse of a {\tt Qubit} within the {\tt Ket} (known as a projection).
This un-entangles the qubit and turns its {\tt Bit} value into {\tt Zero} or {\tt One} (instead of {\tt Unknown} as before the measurement).
Measurement is a probabilistic operation that depends on the amplitude of the current state and its entanglement with other qubits.
If the same quantum circuit followed by measurement is executed several times, it will not in general return the same value due to its probabilistic nature.
However, if repeated sufficiently many times, the actual probability of measuring a 0 (or 1) for a given system can be recovered.
To measure all qubits in a list, we write ``{\tt M >< qs}".

{\tt Reset} is another non-unitary gate that may be used to prepare a qubit in state $\ket{0}$ after it has been measured.  This is a common operation, e.g., in quantum error correction when ancillas are continually measured and then reprepared to be used again. ``{\tt Reset Zero >< qs}" resets all qubits in a list to $\ket{0}$.

Select gates are listed in App.~\ref{app:Gates}. Various functions on {\tt Gate}s exist in \Liquid for easy programming and simulation, including:
\begin{description}
\itemsep 1pt
\parskip 0pt
\item[{\it Gate wrapping}] may be used to wrap a {\tt Gate} definition that may be as simple as a unitary matrix or any number of {\tt Gate}s (sequentially or in parallel) to form a reusable sub-circuit using {\tt WrapOp}.
This makes design and manipulation of quantum algorithms easier.
It allows the programmer to build larger circuits as {\tt Gate}s that contain sub-{\tt Gate}s.
For example, an entire multi-body Hamiltonian term can be implemented as a {\tt Wrap Gate} that might have dozens of primitive gates inside it.
Another example is an adder or Quantum Fourier Transform (QFT) (see Section \ref{sec:Shor}).

\item[{\it Adjoint}] may be used to take the complex conjugate transpose of any unitary {\tt Gate} {\tt U} by writing ``{\tt Adj U}".

\item {\it Reverse} may be used to reverse an entire circuit of unitary gates. It performs the adjoint of all gates along the way and is called by writing ``{\tt let circRev = circ.Reverse()}".
If an {\tt Operation} is implemented as a matrix, {\tt Reverse} may be applied to it.  If a gate is defined as a function or as a non-unitary operation, {\tt Reverse} cannot be applied.

\item {\it Controlled gates} may easily be created using {\tt AddControl}. A single- or multi-qubit unitary {\tt Gate} can be extended into a single- or multi-controlled unitary.  For example, a {\tt CNOT} gate can be built by adding a control to an {\tt X} gate with the command ``{\tt Cgate X qs}".

\item {\it Parametrization} allows dynamic {\tt Gate}s to take any number of parameters as long as the final one is a list of qubits.
For example, consider {\tt Z} rotations by $2\pi i/2^k$ used in the QFT, where $k$ is the parameter (see Sec.~\ref{sec:Prog}).
This can be written in \Liquid as:
\definecolor{mygreen}{rgb}{0,0.3,0}
\definecolor{myred}{rgb}{0.5,0,0}
\lstset{language=Caml,
	basicstyle=\scriptsize,
	tabsize=2,
	morecomment=[l]{//},
    keywordstyle=\color{myred},
    commentstyle=\color{mygreen},
	stringstyle=\color{mygreen},
	showspaces=false,
	showstringspaces=false,
	showtabs=false,
	morekeywords={pown,PI,Cos,Sin}}
\begin{lstlisting}
/// 2pi/2^k gate. 
[<LQD>] let R (k:int) (qs:Qubits) =
	... // Rest of gate definition
	Mat     = (
		let phi     = (2.0*Math.PI)/(pown 2.0 k)
		let phiR    = Math.Cos phi
		let phiI    = Math.Sin phi
		CSMat(2,[(0,0,1.,0.);(1,1,phiR,phiI)]))
 \end{lstlisting}

Non-unitary operations (e.g., {\tt Measure}, {\tt Reset}, {\tt Restore}) may also be parameterized (e.g., reset qubit to $\ket{0}$ or $\ket{1}$).

\item {\it Block operation} enables {\tt Gate}s to be created that operate on a variable number of qubits.
It may be used to operate on subsets of qubits, registers, or entire state vectors.
The only limitation is that all qubits used in a {\tt Gate} must come from a single state vector.
For example, we can apply an {\tt H} gate on all qubits in a list by writing ``{\tt H >< qs}", or alternatively ``{\tt for q in qs do H [q]}" or ``{\tt List.iter (fun q -> H [q]) qs}".
A more detailed example for applying the QFT is given in Section \ref{sec:Prog}.

\item {\it Gate growing} does not affect the algorithm but massively shortens run-times by collapsing sequential unitary gates into a single larger unitary operation.
Trade-offs are made by the system in terms of size of the resulting matrix and density.
There are diminishing returns as the density and size grow; the system optimizes this for best simulation throughput.
%

\item {\it Flatten} turns a hierarchical circuit into a sequence of low-level gates.  This is useful for analysis and resource estimation.

\item {\it Execution} of the circuit is done using {\tt Run}.  Section \ref{sec:Exec} contain details on different modes of execution.

\end{description}

A {\tt Gate} is introspective, so it can ask if it is {\tt Unitary}.
For example, {\tt Adj} requires its operation to be {\tt Unitary} and checks this condition upon the call.
Similarly, a {\tt Gate} can determine if its call parameters match its {\tt Gate} definition, returning an error if there is a mismatch.

The operations that happen behind the scenes on {\tt Qubit}s/{\tt Ket}s require a large amount of complex arithmetic (especially matrix-vector multiples and tensor products).
After working with several native alternatives, we built our own optimized sparse complex linear algebra package in F\# that is highly optimized for this specific application.
Examples include optimized re-use of memory to avoid garbage collection, lazy allocation using skyline vectors based on qubit entanglement, re-ordering of state vectors to turn all tensor products into parallelizable block diagonal operations and many other space and time operations that allow moderate numbers of qubits (30 on a 32GB memory machine) to perform universal quantum operations with no restriction.

At generation time, \Liquid performs the following functions:
Optimization of unitary {\tt Gate}s for efficient Universal simulation (collapsing unitaries together based on size/sparseness);
Optimization of unitary {\tt Gate}s for efficient Hamiltonian simulation (removing non-physical states, exponentiation of the entire circuit…);
Optimization of depth (parallelization of the circuit to compute actual parallel depth);
Replacement of non-available gates (e.g., rotations) to estimate actual depth given a desired substitution method;
Rewriting of Hamiltonian circuits for optimized depth on target hardware (e.g., coalescing of Trotter steps);
Rewriting to map logical to physical qubits with QECC;
Output of circuit to disk as a data structure that could be loaded by other applications;
Output of circuit drawing after any of the above manipulations.

At execution time, it performs:
Function execution for direct simulation of algorithms;
Circuit execution for taking advantage of Generation Time optimization and re-writing;
Injection of user defined unitary and non-unitary noise and statistical analysis;
Debugging for allowing inspection/manipulation of the normally opaque state during execution (one of the benefits of simulation).
\Liquid has the ability to schedule across distributed systems as an ensemble computation in LAN, Cluster and Cloud environments.
The entire system contains more than 30,000 lines of source code.
\Liquid maintains full double-precision complex numbers and in addition re-unitarizes compliled circuit matrices as they drift from unitary due to numerical precision limits.

\subsection{Circuits Manipulators}
A circuit data structure can be passed to a variety of {\tt Circuit} modules, including:
\begin{description}
\itemsep 1pt
\parskip 0pt
\item {\it Decomposition} for replacing unitary gates with low-level gate sequences, primarily to enable fault-tolerant implementation in the laboratory;

\item {\it Optimization} for trading-off circuit depth and width. A simple optimization called {\tt Fold} removes excess identity gates by sliding gates over them (to the left in a circuit diagram) until a non-identity gate is reached;

\item {\it Translation and rule-based rewriting} for mapping to different gate sets and for use in optimization algorithms;

\item {\it Export} for outputting the circuit data structure to a file;

\item {\it Resource costing} for counting the number and types of gates.

\item {\it Quantum error-correcting codes} (QECC) for inserting fault-tolerant protocols for error correction.  Section \ref{sec:QECC} contains an example.

\item {\it Rendering} for drawing a circuit diagram automatically from the \Liquid code.

\end{description}

\subsection{Simulators and Backends}\label{sec:Backend}
Currently, there are two simulators built into the system representing different levels of abstraction: (1) \textit{Universal} for executing a universal quantum computation and (2) \textit{Stabilizer} for efficiently executing a restricted class of gates. \textit{Backends} can be classical machines (for simulation) or an actual quantum computer (for physical implementation).

The \textit{Universal Simulator} is the most flexible of the simulators. It allows a universal set of quantum and classical operations to be performed.  It fully executes the linear algebra and classical control underlying the circuit representation and evolves the full quantum state.  It requires memory resources that grow exponentially with the number of qubits.
It can handle execution of millions of operations (gates), is highly optimized for parallel execution, and is highly efficient in memory usage.
\Liquid has been architected for a virtually unlimited number of qubits (natively 64 bit), but quickly runs out of memory to represent them.
More than a petabyte of main memory would be required to simulate 45 qubits; 32GB of RAM allows roughly 30 qubits.
%

To optimize simulation, we have created classes to embody dynamic arrays that are used as temporary storage throughout the math package that prevent us from both garbage collecting and returning and re-allocating memory that will be used for the same general purpose over time.
All of the storage is globally managed across an entire simulation (instead of on an operation by operation basis).
This ensures an extremely stable memory footprint.

Enhanced parallelization also contributes to the universal simulation speed.
Quantum simulation does not lend itself to distributed computing models.
If the network or disk (no matter how fast they appear to be) need to be accessed, the simulation times will grow to an unacceptable value.
However, efficient use of hardware threads gives massive speed-ups with virtually no cost.
\Liquid has been designed from the ground up with parallelism in mind.
The vast majority of operations are designed to be thread-safe and lock-free.

There is coordination between levels so that hardware threads are not all allocated at a given level if lower levels are able to take better advantage of them.
For example, if a tensor product is executed, it may pay to only use a few threads since the sub-operations are going to be matrix-vector or matrix-matrix multiplies that can better use the available threads for inner-loop operations.

We also make note of the size of the items being worked on and sprout less threads as the work reduces or even take a different code path with no threading when we have reached too small a size for it to be beneficial.
The inner loops have all been optimized for cache coherence and idiosyncrasies of the host language, e.g., 64 bit {\tt for} loops are significantly slower than 32 bit ones in F\# (usually, tail-recursive subroutines are even faster).

We have invested heavily in efficient memory management.
Many of the techniques used are very domain specific.
For example, even though a state vector is typically very large ($2^n$ complex numbers for $n$ qubits), there are many times when only sections of the vector are active (even though it is viewed as dense).
We allocate the vector in blocks (skyline vector) in an on-demand style which allows us to view the entire vector as dense but only lazily created as needed.

A second major savings comes from keeping track of qubit entanglement.
Even though the vector (logically) has $2^n$ entries, if the qubits are all fully un-entangled, we only need to keep $n$ entries.
As entanglement grows (i.e., multi-qubit gates are performed), our memory representation grows.
When measurement occurs on a single entangled qubit, our storage drops by $1/2$.

Measurement is the only simple case where we know that qubits have become dis-entangled.
\Liquid provides an interface for the user to tell the simulator when groups of qubits have become dis-entangled (e.g., at the end of a sub-circuit where registers are no longer entangled).
There is also a version of this call that actually checks if the qubits are really unentangled (very expensive) that helps the user check assertions of her circuit.

We have also developed a package on top of the universal simulator that provides simulation of Hamiltonians.
The simulation environment attempts to model some of the realistic physics in a quantum system developed in a laboratory.
It differs from the other simulators in that it has the concept of the time it takes for an operation to be performed (since it is numerically solving a differential equation).
It is also (by its very nature) slow due to the requirements for simulating a state evolving over time.
An example Hamiltonian simulation is given in Appendix \ref{app:Ham}.

The \textit{Stabilizer Simulator} is a restricted simulator based on methods in Ref.~\cite{Aaronson2004}.
It performs a specialized class of quantum operations (the so-called ``Clifford-group" operations).
It evolves only the stabilizer information in a matrix tableau, rather than the full quantum state.
Thus, it requires memory resources that grow linearly with the number of qubits.  The set of circuits simulable includes most quantum error correction protocols.
Efficient simulation offerings could be extended to include methods in Refs.~\cite{Garcia2013,Viamontes2005}.

The Stabilizer simulator has the virtue of allowing large circuits (millions of operations) on massive numbers of qubits (tens of thousands).
The main limitation is the types of gates which may be included in the circuit. They are fixed in the system and come from the “stabilizer” class (e.g., Clifford group).
This limits the usefulness of the types of algorithms that can be implemented and tested.
However, it does allow the design and test of Quantum Error Correction Codes (QECC) which requires large numbers of physical qubits per logical qubits.
An example usage of the Stabilizer simulator is given in Section \ref{sec:Prog}.

\subsection{Execution Modes}\label{sec:Exec}
\Liquid code can be executed in several ways:
\begin{enumerate}
\itemsep 1pt
\parskip 0pt
\item {\em Test mode}: Many built-in tests of the system can be invoked from the command line and are useful demonstrations, including all examples provided in this paper (see App.~\ref{app:Tests}).
\item {\em Script mode}: The system can be run directly from an F\# text script (.fsx file). This allows the simulator to be operated by simply running the executable (no separate language compilation required). The entire simulator is available from this mode, but interactive debugging is difficult and start-up times are slower. Script mode allows users to experiment with fast turn-around time and ease of use (no need to install a complete development environment).
This is also the method used for submission to Cloud services.
\item {\em Function mode}:  This is the normal development mode. It requires a compilation environment (e.g., Visual Studio) and the use of a .Net language (typically F\#). The user has the full range of APIs at her disposal and can extend the environment in many ways as well as building her own complete applications.
Here is the actual top level of the \Liquid executable:
\definecolor{mygreen}{rgb}{0,0.3,0}
\definecolor{myred}{rgb}{0.5,0,0}
\lstset{language=Caml,
	  basicstyle=\scriptsize,
	  tabsize=2,
	  morecomment=[l]{//},
      keywordstyle=\color{myred},
      commentstyle=\color{mygreen},
	stringstyle=\color{mygreen},
	showspaces=false,
	showstringspaces=false,
	showtabs=false,
	morekeywords={Parser,EntryPoint,GetCommandLineArgs,CommandArgs,toList,skip,CommandRun}}
\begin{lstlisting}
    [<EntryPoint>]
    let Main _ =
        let args    = // Skip the program name
            Environment.GetCommandLineArgs()
            |> Seq.skip 1 |> Seq.toList
        let p       = Parser(args)
        let las     = p.CommandArgs()
        p.CommandRun las // Run the command line
\end{lstlisting}
A user may implement this, mark any callable functions with the {\tt[<LQD>]} attribute and then link with the \Liquid libraries. Then the user can write: ``{\tt Liquid UserFunc(args,\ldots)}" and get all the command line features built into the parser.

\item {\em Circuit mode}: Function mode can be compiled into a circuit data structure on qubits {\tt qs} that can be simulated with ``{\tt circ.Run qs}".
This data structure can be manipulated by the user, run through built-in optimizers, have quantum error correction added, rendered as drawings, exported for use in other environments, and may be run directly by all the simulation engines.

\end{enumerate}

\subsection{Environments}\label{sec:Env}
The two ways to interact with the system are via a full compilation environment in Visual Studio linked to the \Liquid library (dll), or via an F\# script hosted by the \Liquid application (exe).
Both provide advantages. Compilation provides IntelliSense editing and a full debugging environment, while scripting provides a quick and easy way to prototype and extend \Liquid while quickly turning around simulations with varying parameters.

Any function in the system that is tagged with the {\tt [<LQD>]} attribute may be called from the command line (including any user extensions). For example the function {\tt showStr(<string>)} will show a string on the console. This function is marked with {\tt [<LQD>]} and can be invoked directly:
\begin{alltt}\scriptsize
    \emph{> Liquid __show(``Hello world")}
    Hello world
\end{alltt}
Some very sophisticated functions are built into the system and are demonstrated in the example for running Shor's algorithm (Sec.~\ref{sec:Shor}).

\Liquid also has the ability to run in a fully distributed manner via ensemble computations.
Often, simulations of quantum circuits are run a large number of times with either slightly different circuits or parameters or to check statistical results.
Ensemble computations are accomplished easily by defining an {\tt Ensemble.xml} file.
An example ensemble run on 5 machines is written as:
\definecolor{mygreen}{rgb}{0,0.3,0}
\definecolor{myred}{rgb}{0.5,0,0}
\lstset{language=XML,
      basicstyle=\scriptsize,
      keywordstyle=\color{myred},
      commentstyle=\color{green},
	stringstyle=\color{mygreen},
	tabsize=2,
	morekeywords={Pars,Exe,Host,Cmd,Shor,Ensemble}}
\begin{lstlisting}
<Ensemble Default="Shor">
  <Pars>
    <Exe>\\machine00\Liquid\Liquid.exe</Exe>
    <Host>machine00</Host>
    <Host>machine01</Host>
    <Host>machine02</Host>
    <Host>machine03</Host>
    <Host>machine04</Host>
  </Pars>
  <Shor Count="12" Args='/pfx "%N4%|"'>
    <Cmd Range="1,1,2" Name="129">__Shor(%N%,true)</Cmd>
    <Cmd Range="1,1,2" Name="259">__Shor(%N%,true)</Cmd>
    <Cmd Range="1,1,2" Name="513">__Shor(%N%,true)</Cmd>
    <Cmd Range="1,1,2" Name="1025">__Shor(%N%,true)</Cmd>
    <Cmd Range="1,1,2" Name="2049">__Shor(%N%,true)</Cmd>
    <Cmd Range="1,1,2" Name="4097">__Shor(%N%,true)</Cmd>
  </Shor>
</Ensemble>
\end{lstlisting}

We define the command {\tt Shor} which will factor 6 numbers twice ({\tt Count="12"}) across the machines.
\Liquid did not have to be installed on any of the other machines.
When the ensemble command is given to \Liquid, it will install itself as a Windows Service on all of the other machines, start them up, run the simulations, and then shut down the services. All of this is invisible to the user.


\section{Code Example: Quantum Teleportation}\label{sec:Prog}

\subsection{The Circuit}
We now present the \Liquid code for quantum teleportation:
\definecolor{mygreen}{rgb}{0,0.3,0}
\definecolor{myred}{rgb}{0.5,0,0}
\lstset{language=Caml,
	  basicstyle=\scriptsize,
	  tabsize=2,
	  morecomment=[l]{//},
      keywordstyle=\color{myred},
      commentstyle=\color{mygreen},
	stringstyle=\color{mygreen},
	showspaces=false,
	showstringspaces=false,
	showtabs=false,
	morekeywords={EPR,H,CNOT,teleport,LabelL,M,BC,LabelR}}
\begin{lstlisting}
// Define an EPR function
let EPR (qs:Qubits) =  H qs; CNOT qs

// Teleport qubit 0 to qubit 2
let teleport (qs:Qubits) =
  let qs' = qs.Tail

  LabelL >!< // Give names to the qubits
    (["Src";"\\ket{0}";"\\ket{0}"],qs)

  EPR qs'; CNOT qs; H qs
  M qs'; BC X qs'					// Maybe apply X
  M qs; BC Z !!(qs,0,2)		// Maybe apply Z
  LabelR "Dest" !!(qs,2)	// Label output
\end{lstlisting}

We define a function called {\tt EPR} that takes a list of qubits and then applies a Hadamard gate to the first qubit and a {\tt CNOT} to the first two qubits.
By convention, gates will take as many qubits as they require from the beginning of the list.
If a gate can take a variable number of qubits (like a quantum Fourier Transform) then a list of the length to be used must be provided.


Now we can use the EPR function within a {\tt teleport} function.
In the first line of the function we take the {\tt Tail} of the qubit list so that we are left with qubits 1 and 2 (named {\tt qs'}).
Now we label all the qubits with names for drawing. {\tt LabelL} is an example of a non-unitary gate that puts information in any renderings of the circuit, but does not affect the circuit simulation in any way.
The {\tt >!<} function is an example of a \Liquid specific operator that maps a gate to a list of arguments.
Now we call the EPR function previously defined.
We then perform a {\tt CNOT} and {\tt H} on the first two qubits.
To receive, the message we measure qubit 1 and conditionally apply an {\tt X} gate to qubit 2 depending on the value measured.
This binary control gate ({\tt BC}) is another example of a non-unitary gate. We then repeat with {\tt Z} gate on qubit 2, controlled by qubit 0. Finally we place a drawing {\tt Label} on qubit 2.

With the {\tt teleport} \Liquid function, we can perform several operations as depicted below:
%
\definecolor{mygreen}{rgb}{0,0.3,0}
\definecolor{myred}{rgb}{0.5,0,0}
\lstset{language=Caml,
	basicstyle=\scriptsize,
	tabsize=2,
	morecomment=[l]{//},
    keywordstyle=\color{myred},
    commentstyle=\color{mygreen},
	stringstyle=\color{mygreen},
	showspaces=false,
	showstringspaces=false,
	showtabs=false,
	morekeywords={EPR,H,CNOT,teleport,LabelL,M,BC,LabelR,Ket,Circuit,Run,Dump,Compile,GrowGates,Fold,RenderHT}}
\begin{lstlisting}
let ket	= Ket(3)		// Create state
let qs	= ket.Qubits
teleport qs					// Run Teleport
let circ 	= 				// Compile to circuit
    Circuit.Compile teleport qs
circ.Run qs					// Run circuit
circ.Dump()					// Dump gates to log
circ.Fold()					// Fold the circuit
    .RenderHT(``Teleport'')	// Draw HTML and TeX
let circ2	= 				// Grow Unitaries together
    circ.GrowGates ket
circ2.Run qs				// Run the optimized circuit
\end{lstlisting}

%
To begin, we create a state vector ({\tt Ket}) of 3 qubits and get a reference to those qubits ({\tt qs}).
The line {\tt teleport qs} calls {\tt teleport} and runs it on the state vector.
We can map {\tt teleport} into a {\tt Circuit} data structure by compiling it into {\tt circ}.
This can be run before or after any manipulations.
The {\tt Dump} command
provides complete information about the item begin exported.
In this case, we get a complete specification for all parts of the teleport circuit (part is shown below):
%
\definecolor{mygreen}{rgb}{0,0.3,0}
\definecolor{myred}{rgb}{0.5,0,0}
\lstset{language=Caml,
	basicstyle=\scriptsize,
	tabsize=2,
	morecomment=[l]{//},
    keywordstyle=\color{myred},
    commentstyle=\color{mygreen},
	stringstyle=\color{mygreen},
	showspaces=false,
	showstringspaces=false,
	showtabs=false}
\begin{lstlisting}
SEQ
  APPLY
    GATE H is a Hadamard
      0.7071 0.7071
      0.7071 -0.7071
    WIRE(Id:1)
    WIRE(Id:2)
  APPLY
    GATE CNOT is a Controlled NOT
      1 0 0 0
      0 1 0 0
      0 0 0 1
      0 0 1 0
    WIRE(Id:1)
    WIRE(Id:2)
  PAR
    APPLY
      GATE Meas is a Collapse State
        1 0
        0 1
      WIRE(Id:1)
  BitCon
    GATE BitControl
    WIRE(Id:1)
    WIRE(Id:2)
      APPLY
        GATE X is a Pauli X flip
          0 1
          1 0
        WIRE(Id:2)
\end{lstlisting}

%
We see a {\tt SEQ}uence of gate {\tt APP}lications (here the {\tt CNOT} and {\tt H} gates after {\tt EPR}) followed by the first of the Binary Control ({\tt BC}) gates.

The two operations {\tt GrowGates} and {\tt Run}
optimize the circuit by growing gates into larger unitaries and then runs the optimized circuit.
Now the sources for {\tt teleport} ({\tt EPR}, {\tt CNOT}, and {\tt H}) have been combined into one gate as:
%
\definecolor{mygreen}{rgb}{0,0.3,0}
\definecolor{myred}{rgb}{0.5,0,0}
\lstset{language=Caml,
	basicstyle=\scriptsize,
	tabsize=2,
	morecomment=[l]{//},
    keywordstyle=\color{myred},
    commentstyle=\color{mygreen},
	stringstyle=\color{mygreen},
	showspaces=false,
	showstringspaces=false,
	showtabs=false}
\begin{lstlisting}
SEQ
  APPLY
    GATE 64B8DB5 is a grown gate
      0.5 0 0.5 0 0 0.5 0 -0.5
      0 0.5 0 0.5 0.5 0 -0.5 0
      0 0.5 0 -0.5 0.5 0 0.5 0
      0.5 0 -0.5 0 0 0.5 0 0.5
      0.5 0 0.5 0 0 -0.5 0 0.5
      0 0.5 0 0.5 -0.5 0 0.5 0
      0 0.5 0 -0.5 -0.5 0 -0.5 0
      0.5 0 -0.5 0 0 -0.5 0 -0.5
    WIRE(Id:0)
    WIRE(Id:1)
    WIRE(Id:2)
\end{lstlisting}


Finally, the circuit can be parallelized by removing identity gates ({\tt Fold()}) and then {\tt Render}ed to a file as shown in Fig.~\ref{fig:TeleDraw}.
The rendering contains all of the elements we defined, plus information about the qubits (via double wires) showing where they were converted to binary ({\tt Bit}) values after being measured.
{\tt RenderHT} generates both HTML (SVG graphics) and \TeX (TIKZ) files, as shown in this paper.
Note that the examples throughout use only destructive measurement, however non-destructive measurements are also available in \LiquidB.

\subsection{The Circuit with Error Correction}\label{sec:QECC}

A necessary step in targeting a high-level representation of a quantum algorithm to a low-level quantum hardware architecture is the insertion of quantum error correction circuitry (see \cite{Nielsen2000} for review of quantum error correction).
The use of quantum error correction can help reduce the probability of errors in a given quantum circuit by replacing it with a fault-tolerant, noise-reducing circuit.
Each \emph{logical} qubit is encoded in a set of \emph{physical} qubits using a quantum error correction circuit.  The exact circuit depends on the particular quantum code being used.
Similarly, a logical gate is replaced by an encoded circuit operating at the level of physical gates.
An encoded computation thus requires substantially more resources than an unencoded computation, but when the components operate below a certain error threshold, it reduces the probability of errors at the logical level of computation.
To enable investigation of quantum error-correcting codes (QECC), \Liquid includes packages to replace logical gates and qubits with error correction protocols involving physical qubits and gates.

As an example, consider the $[[7,1,3]]$ Steane code  (see \cite{Nielsen2000} for details) which encodes a single logical qubit in $7$ physical qubits and can correct one physical error.
To encode the {\tt teleport} function, we may write:
%
\definecolor{mygreen}{rgb}{0,0.3,0}
\definecolor{myred}{rgb}{0.5,0,0}
\lstset{language=Caml,
	  basicstyle=\scriptsize,
	  tabsize=3,
	  morecomment=[l]{//},
      keywordstyle=\color{myred},
      commentstyle=\color{mygreen},
	stringstyle=\color{mygreen},
	showspaces=false,
	showstringspaces=false,
	showtabs=false,
	morekeywords={X,teleport,M,Compile,tele1,Steane7,Inject,Stabilizer,Run,Log2Phys,Decode}}
\begin{lstlisting}
let tele1 (qs:Qubits) =		// Stabilizer friendly teleport
	X qs							// teleport a |1>
	teleport qs					// do the circuit		
	!!(qs,2)						// measure at the end
let circ		= Circuit.Compile tele1 qs
let s7		= Steane7(circ)			// Apply a Steane code
let errC,stats	= s7.Inject 0.01		// Inject errors
let stab		= Stabilizer(errC,ket)	// Setup simulation
stab.Run()									// Run the simulation

// Convert physical result to logical result
let bit0,dist0	= s7.Log2Phys 0 |> s7.Decode
let bit1,dist1	= s7.Log2Phys 1 |> s7.Decode
let bit2,dist2	= s7.Log2Phys 2 |> s7.Decode
\end{lstlisting}

%
Here we have wrapped the {\tt teleport} function in a new function ({\tt tele1}) which flips the message qubit (prepares a $\ket{1}$), teleports it, and then measures the result.
First we compile this function into a circuit and instantiate one of the QECC classes ({\tt Steane7}) which transforms the circuit from the logical level to the physical level by encoding each logical qubit in 7 physical qubits.
Each logical gate is also replaced with physical-level gates.

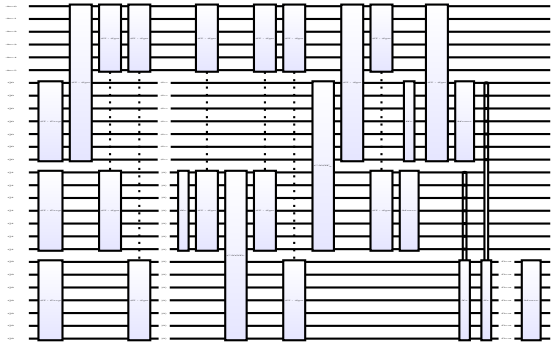
\begin{figure}[h]
\centering
\begin{tikzpicture}[scale=1.00,every node/.style={xscale=0.2,yscale=0.07}]
\matrix[matStyle] {
  \lbl[0-0]{Anc0} & \point[0-1] &  & \mg[0-3]{S7:Syn} & \mg[0-4]{S7:Syn} & \mg[0-5]{S7:Syn} &  &  & \mg[0-8]{S7:Syn} &  & \mg[0-10]{S7:Syn} & \mg[0-11]{S7:Syn} &  & \mg[0-13]{S7:Syn} & \mg[0-14]{S7:Syn} &  & \mg[0-16]{S7:Syn} &  &  &  &  & \point[0-21]  \\
  \lbl[1-0]{Anc1} & \point[1-1] &  &  &  &  &  &  &  &  &  &  &  &  &  &  &  &  &  &  &  & \point[1-21]  \\
  \lbl[2-0]{Anc2} & \point[2-1] &  &  &  &  &  &  &  &  &  &  &  &  &  &  &  &  &  &  &  & \point[2-21]  \\
  \lbl[3-0]{Anc3} & \point[3-1] &  &  &  &  &  &  &  &  &  &  &  &  &  &  &  &  &  &  &  & \point[3-21]  \\
  \lbl[4-0]{Anc4} & \point[4-1] &  &  &  &  &  &  &  &  &  &  &  &  &  &  &  &  &  &  &  & \point[4-21]  \\
  \lbl[5-0]{Anc5} & \point[5-1] &  &  & \mg[5-4]{S7:Syn} & \mg[5-5]{S7:Syn} &  &  & \mg[5-8]{S7:Syn} &  & \mg[5-10]{S7:Syn} & \mg[5-11]{S7:Syn} &  &  & \mg[5-14]{S7:Syn} &  &  &  &  &  &  & \point[5-21]  \\
  \lbl[6-0]{Q0} & \point[6-1] & \mg[6-2]{S7:Prep} &  &  &  & \lbl[6-6]{Src} &  &  &  &  &  & \mg[6-12]{CNOT_T} &  &  & \mg[6-15]{H_T} &  & \mg[6-17]{Meas_T} & \ctrl[6-18] &  &  & \point[6-21]  \\
  \lbl[7-0]{Q0} & \point[7-1] &  &  &  &  & \lbl[7-6]{Src} &  &  &  &  &  &  &  &  &  &  &  & \ctrl[7-18] &  &  & \point[7-21]  \\
  \lbl[8-0]{Q0} & \point[8-1] &  &  &  &  & \lbl[8-6]{Src} &  &  &  &  &  &  &  &  &  &  &  & \ctrl[8-18] &  &  & \point[8-21]  \\
  \lbl[9-0]{Q0} & \point[9-1] &  &  &  &  & \lbl[9-6]{Src} &  &  &  &  &  &  &  &  &  &  &  & \ctrl[9-18] &  &  & \point[9-21]  \\
  \lbl[10-0]{Q0} & \point[10-1] &  &  &  &  & \lbl[10-6]{Src} &  &  &  &  &  &  &  &  &  &  &  & \ctrl[10-18] &  &  & \point[10-21]  \\
  \lbl[11-0]{Q0} & \point[11-1] &  &  &  &  & \lbl[11-6]{Src} &  &  &  &  &  &  &  &  &  &  &  & \ctrl[11-18] &  &  & \point[11-21]  \\
  \lbl[12-0]{Q0} & \point[12-1] & \mg[12-2]{S7:Prep} & \mg[12-3]{S7:Syn} &  &  & \lbl[12-6]{Src} &  &  &  &  &  &  & \mg[12-13]{S7:Syn} &  & \mg[12-15]{H_T} & \mg[12-16]{S7:Syn} & \mg[12-17]{Meas_T} & \ctrl[12-18] &  &  & \point[12-21]  \\
  \lbl[13-0]{Q1} & \point[13-1] & \mg[13-2]{S7:Prep} &  & \mg[13-4]{S7:Syn} &  & \lbl[13-6]{\ket{0}} & \mg[13-7]{H_T} & \mg[13-8]{S7:Syn} & \mg[13-9]{CNOT_T} & \mg[13-10]{S7:Syn} &  &  &  & \mg[13-14]{S7:Syn} & \mg[13-15]{Meas_T} &  & \ctrl[13-17] &  &  &  & \point[13-21]  \\
  \lbl[14-0]{Q1} & \point[14-1] &  &  &  &  & \lbl[14-6]{\ket{0}} &  &  &  &  &  &  &  &  &  &  & \ctrl[14-17] &  &  &  & \point[14-21]  \\
  \lbl[15-0]{Q1} & \point[15-1] &  &  &  &  & \lbl[15-6]{\ket{0}} &  &  &  &  &  &  &  &  &  &  & \ctrl[15-17] &  &  &  & \point[15-21]  \\
  \lbl[16-0]{Q1} & \point[16-1] &  &  &  &  & \lbl[16-6]{\ket{0}} &  &  &  &  &  &  &  &  &  &  & \ctrl[16-17] &  &  &  & \point[16-21]  \\
  \lbl[17-0]{Q1} & \point[17-1] &  &  &  &  & \lbl[17-6]{\ket{0}} &  &  &  &  &  &  &  &  &  &  & \ctrl[17-17] &  &  &  & \point[17-21]  \\
  \lbl[18-0]{Q1} & \point[18-1] &  &  &  &  & \lbl[18-6]{\ket{0}} &  &  &  &  &  &  &  &  &  &  & \ctrl[18-17] &  &  &  & \point[18-21]  \\
  \lbl[19-0]{Q1} & \point[19-1] & \mg[19-2]{S7:Prep} &  & \mg[19-4]{S7:Syn} &  & \lbl[19-6]{\ket{0}} & \mg[19-7]{H_T} & \mg[19-8]{S7:Syn} &  & \mg[19-10]{S7:Syn} &  & \mg[19-12]{CNOT_T} &  & \mg[19-14]{S7:Syn} & \mg[19-15]{Meas_T} &  & \ctrl[19-17] &  &  &  & \point[19-21]  \\
  \lbl[20-0]{Q2} & \point[20-1] & \mg[20-2]{S7:Prep} &  &  & \mg[20-5]{S7:Syn} & \lbl[20-6]{\ket{0}} &  &  &  &  & \mg[20-11]{S7:Syn} &  &  &  &  &  & \mg[20-17]{X_T} & \mg[20-18]{Z_T} & \lbl[20-19]{Dest} & \mg[20-20]{Meas_T} & \point[20-21]  \\
  \lbl[21-0]{Q2} & \point[21-1] &  &  &  &  & \lbl[21-6]{\ket{0}} &  &  &  &  &  &  &  &  &  &  &  &  & \lbl[21-19]{Dest} &  & \point[21-21]  \\
  \lbl[22-0]{Q2} & \point[22-1] &  &  &  &  & \lbl[22-6]{\ket{0}} &  &  &  &  &  &  &  &  &  &  &  &  & \lbl[22-19]{Dest} &  & \point[22-21]  \\
  \lbl[23-0]{Q2} & \point[23-1] &  &  &  &  & \lbl[23-6]{\ket{0}} &  &  &  &  &  &  &  &  &  &  &  &  & \lbl[23-19]{Dest} &  & \point[23-21]  \\
  \lbl[24-0]{Q2} & \point[24-1] &  &  &  &  & \lbl[24-6]{\ket{0}} &  &  &  &  &  &  &  &  &  &  &  &  & \lbl[24-19]{Dest} &  & \point[24-21]  \\
  \lbl[25-0]{Q2} & \point[25-1] &  &  &  &  & \lbl[25-6]{\ket{0}} &  &  &  &  &  &  &  &  &  &  &  &  & \lbl[25-19]{Dest} &  & \point[25-21]  \\
  \lbl[26-0]{Q2} & \point[26-1] & \mg[26-2]{S7:Prep} &  &  & \mg[26-5]{S7:Syn} & \lbl[26-6]{\ket{0}} &  &  & \mg[26-9]{CNOT_T} &  & \mg[26-11]{S7:Syn} &  &  &  &  &  & \mg[26-17]{X_T} & \mg[26-18]{Z_T} & \lbl[26-19]{Dest} & \mg[26-20]{Meas_T} & \point[26-21]  \\
   &  &  &  &  &  &  &  &  &  &  &  &  &  &  &  &  &  &  &  &  &   \\
};
\multigate{0-3}{12-3}{S7:Syn}
\multigate{0-4}{5-4}{S7:Syn}
\multigate{0-5}{5-5}{S7:Syn}
\multigate{0-8}{5-8}{S7:Syn}
\multigate{0-10}{5-10}{S7:Syn}
\multigate{0-11}{5-11}{S7:Syn}
\multigate{0-13}{12-13}{S7:Syn}
\multigate{0-14}{5-14}{S7:Syn}
\multigate{0-16}{12-16}{S7:Syn}
\multigate{6-2}{12-2}{S7:Prep}
\multigate{6-12}{19-12}{CNOT_T}
\multigate{6-15}{12-15}{H_T}
\multigate{6-17}{12-17}{Meas_T}
\multigate{13-2}{19-2}{S7:Prep}
\multigate{13-4}{19-4}{S7:Syn}
\multigate{13-7}{19-7}{H_T}
\multigate{13-8}{19-8}{S7:Syn}
\multigate{13-9}{26-9}{CNOT_T}
\multigate{13-10}{19-10}{S7:Syn}
\multigate{13-14}{19-14}{S7:Syn}
\multigate{13-15}{19-15}{Meas_T}
\multigate{20-2}{26-2}{S7:Prep}
\multigate{20-5}{26-5}{S7:Syn}
\multigate{20-11}{26-11}{S7:Syn}
\multigate{20-17}{26-17}{X_T}
\multigate{20-18}{26-18}{Z_T}
\multigate{20-20}{26-20}{Meas_T}
\begin{pgfonlayer}{background}
\qw{0-1}{0-21} \qw{1-1}{1-21} \qw{2-1}{2-21} \qw{3-1}{3-21} \qw{4-1}{4-21} \qw{5-1}{5-21}
\qw{6-1}{6-21} \qw{7-1}{7-21} \qw{8-1}{8-21} \qw{9-1}{9-21} \qw{10-1}{10-21} \qw{11-1}{11-21}
\qw{12-1}{12-21} \qw{13-1}{13-21} \qw{14-1}{14-21} \qw{15-1}{15-21} \qw{16-1}{16-21}
\qw{17-1}{17-21} \qw{18-1}{18-21} \qw{19-1}{19-21} \qw{20-1}{20-21} \qw{21-1}{21-21}
\qw{22-1}{22-21} \qw{23-1}{23-21} \qw{24-1}{24-21} \qw{25-1}{25-21} \qw{26-1}{26-21}
\cwx{6-18}{26-18} \cwx{13-17}{26-17} \dwx{5-4}{13-4} \dwx{5-8}{13-8} \dwx{5-10}{13-10}
\dwx{5-14}{13-14} \dwx{5-5}{20-5} \dwx{5-11}{20-11}
\end{pgfonlayer}
\end{tikzpicture}
\caption{\label{fig:S7}High-level view of {\tt teleport} after the addition of quantum error correction using the {\tt QECC} class.}
\end{figure}

\begin{figure*}[hbt]
\centering
\includegraphics[width=7in]{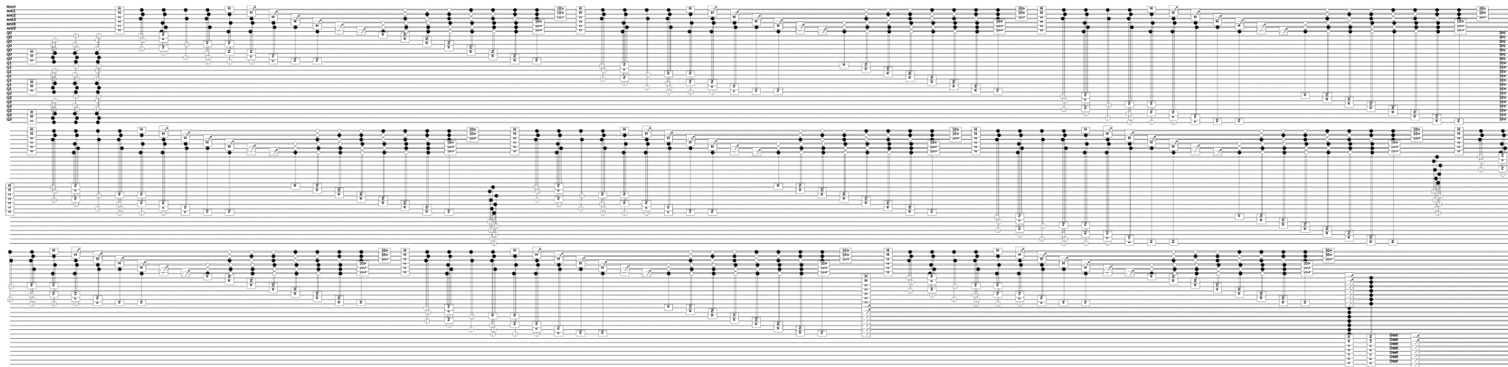}
\caption{\label{fig:S7detail}Detailed view of {\tt teleport} after QECC.}
\end{figure*}

The {\tt Steane7} class is derived from the abstract {\tt QECC} class.
The {\tt QECC} class can be easily extended by the user to permit other codes such as concatenated codes and topological codes like the surface code.
The circuit created ({\tt s7}) contains many more qubits and gates than the original logical-level {\tt teleport} circuit.
A high-level view of ({\tt s7}) is shown in Fig.~\ref{fig:S7}.
Here, the boxes represent parts of the QEC routine, such as encoding, syndrome preparation, syndrome extraction, and correction.
Fig.~\ref{fig:S7detail} shows {\tt s7} at the level of physical qubits and operations.
The three logical qubits are encoded in $21$ physical qubits. The other qubits shown are ancilla qubits used for error syndrome extraction.
In this example, we have chosen not to apply error correction to idle circuit locations (identity gates).

Figures \ref{fig:S7} and \ref{fig:S7detail} show quantum error correction layered over the quantum teleportation circuit at different levels of detail. \Liquid allows drawing circuits at different levels of abstraction, depending on the needs of the user. For example, Fig.~\ref{fig:S7} is useful when examining qubit usage and parallelization, while Fig.~\ref{fig:S7detail} is useful for verifying the circuit in its entirety. Both levels are of great use to algorithm developers.

The challenge in simulating large quantum error-correcting codes on a classical computer is that we quickly run out of qubits since each logical qubit is encoded in a few to thousands of physical qubits depending on the code.
There is a better solution. We can switch from the Universal to the Stabilizer simulator.
This is what the call above
to {\tt Stabilizer()} does.
Say, for example, that we are interested in modeling errors. We may first {\tt Inject} depolarizing errors ({\tt X}, {\tt Y}, or {\tt Z} gates) with a given probability to create an error circuit ({\tt errC}) and then create an instance of the Stabilizer simulator to run.
\Liquid can easily handle simulations of tens of thousands of qubits in this way.
The last three lines in the code above
convert (decode) physical qubits back to logical ones so we can check if we teleported the proper message.
The distances between the encoded logical qubits and the expected codewords are also returned.
More realistic noise models that involve non-unitary operations (see \cite{Nielsen2000} for examples) can be modeled using the Universal simulator.

\section{Shor's Algorithm in \LiquidT}\label{sec:Shor}

Quantum algorithms find solutions to some problems exponentially faster than the corresponding best-known classical algorithms.
The most famous example is Shor's polynomial-time quantum algorithm for prime factorization \cite{Shor1994}.
The algorithm uses an important primitive called the quantum Fourier transform (QFT).
It also requires classical pre- and post-processing and quantum circuits for modular arithmetic.

\begin{figure}[h]
\centering
\begin{tikzpicture}[scale=1.00,every node/.style={scale=0.80}]
\matrix[matStyle] {
\lbl[0-0]{\ket{0}} & \point[0-1] & \gate[0-2]{H^{\otimes t}} & \ctrl[0-3] & \gate[0-4]{QFT^\dagger} & \meter[0-5] & \point[0-6] \\
\lbl[1-0]{\ket{1}} & \point[1-1] & & \gate[1-3]{a^{j} mod N} & & & \point[1-6] \\
& & & & & & \\
};
\begin{pgfonlayer}{background}
\qw{0-1}{0-5} \qw{1-1}{1-6} \cw{0-5}{0-6} \qwx{0-3}{1-3};
\draw[thick] (0-1) ++(0,-.15) -- ++(70:.3); 
\draw[thick] (1-1) ++(0,-.15) -- ++(70:.3); 
\path (0-3) ++(-.6,.3) node {\tiny $\ket{j}$};
\end{pgfonlayer}
\end{tikzpicture}
\caption{\label{fig:OrderFind}High-level circuit for order finding \cite{Nielsen2000}.}
\end{figure}
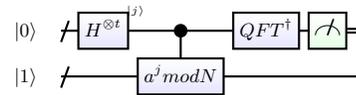

At a high level, Shor's algorithm begins with classical pre-processing of the $n$-bit number $N$ to be factored.
At the heart of the algorithm is quantum order finding, which determines the least positive integer $r$ such that $a^r \mod N$ is congruent to $1$.
It is shown at a high level in Fig.~\ref{fig:OrderFind} and executes as follows:
a register of quantum states is placed in superposition and a second register of quantum states is initialized to $\ket{1}$.
\footnote{The second register is initialized to $\ket{1}$ for simplicity since at the start of the algorithm the order $r$ is unknown making it impossible to prepare the eigenstates of powers of $a^r\mod N$.
However, conveniently $\frac{1}{\sqrt{r}}\sum_{s=0}^{r-1} \ket{u_s} = \ket{1}$.
}
Next a controlled application of modular exponentiation is applied (modular $N$) between two quantum registers, followed by an inverse QFT applied to the top quantum register.
Finally, classical post-processing is performed to find the factors, or the algorithm is repeated if none are found.

\subsection{Code Example: Order Finding}
Quantum order finding requires a quantum oracle to perform modular exponentiation and a quantum Fourier transform (we follow the circuit given in \cite{Nielsen2000}).
Here, we present circuit examples for these routines and the corresponding \Liquid code.

\textbf{Quantum Fourier Transform. }
The QFT is an important primitive that can be performed using only $O(n^2)$ quantum operations, in contrast to $\Theta(n 2^n)$ classical operations for a discrete Fourier transform.
It may also be used, for example, within quantum phase estimation and quantum arithmetic functions.

The \Liquid code for the inverse QFT ({\tt QFT')} on an arbitrary number of qubits is given by:
\definecolor{mygreen}{rgb}{0,0.3,0}
\definecolor{myred}{rgb}{0.5,0,0}
\lstset{language=Caml,
	  basicstyle=\scriptsize,
	  tabsize=2,
	  morecomment=[l]{//},
      keywordstyle=\color{myred},
      commentstyle=\color{mygreen},
	stringstyle=\color{mygreen},
	showspaces=false,
	showstringspaces=false,
	showtabs=false,
	morekeywords={CR,H}}
\begin{lstlisting}
let QFT' (qs:Qubits) =
	let n   = qs.Length						// Get number of qubits
	for aIdx in 0..n-1 do					// Process each qubit
		let a   = qs.[aIdx]					// Get the current qubit
		for k in aIdx+1..-1..2 do		// Walk each control qubit
			let c   = qs.[aIdx-(k-1)]	// Extract the control
			CR' k [c;a]					// Apply the controlled rotation
		H [a]									// Hadamard each when done
\end{lstlisting}

The corresponding diagram generated by the \Liquid source code applied to 5 qubits is shown in Fig.~\ref{fig:QFT-5}.
Note the use of controlled adjoint rotations ({\tt CR'}) which uses the {\tt Cgate}, {\tt Adj}, and {\tt R} definitions described earlier.

\begin{figure}[h]
\centering
\begin{tikzpicture}[scale=1.00,every node/.style={scale=0.50}]
\matrix[matStyle] {
\lbl[5-0]{\ket{b0}} & \point[5-1] & \gate[5-2]{H} & \ctrl[5-3] & & \ctrl[5-5] & & & \ctrl[5-8] & & & & \ctrl[5-12] & & & & & \point[5-17] \\
\lbl[6-0]{\ket{b1}} & \point[6-1] & & \gate[6-3]{R2^\dagger} & \gate[6-4]{H} & & \ctrl[6-6] & & & \ctrl[6-9] & & & & \ctrl[6-13] & & & & \point[6-17] \\
\lbl[7-0]{\ket{b2}} & \point[7-1] & & & & \gate[7-5]{R3^\dagger} & \gate[7-6]{R2^\dagger} & \gate[7-7]{H} & & & \ctrl[7-10] & & & & \ctrl[7-14] & & & \point[7-17] \\
\lbl[8-0]{\ket{b3}} & \point[8-1] & & & & & & & \gate[8-8]{R4^\dagger} & \gate[8-9]{R3^\dagger} & \gate[8-10]{R2^\dagger} & \gate[8-11]{H} & & & & \ctrl[8-15] & & \point[8-17] \\
\lbl[9-0]{\ket{b4}} & \point[9-1] & & & & & & & & & & & \gate[9-12]{R5^\dagger} & \gate[9-13]{R4^\dagger} & \gate[9-14]{R3^\dagger} & \gate[9-15]{R2^\dagger} & \gate[9-16]{H} & \point[9-17] \\
& & & & & & & & & & & & & & & & & \\
};
\begin{pgfonlayer}{background}
\qw{5-1}{5-17}
\qw{6-1}{6-17} \qw{7-1}{7-17} \qw{8-1}{8-17} \qw{9-1}{9-17} \qwx{5-3}{6-3} \qwx{5-5}{7-5}
\qwx{5-8}{8-8} \qwx{5-12}{9-12} \qwx{6-6}{7-6} \qwx{6-9}{8-9} \qwx{6-13}{9-13} \qwx{7-10}{8-10}
\qwx{7-14}{9-14} \qwx{8-15}{9-15}
\end{pgfonlayer}
\end{tikzpicture}
\caption{\label{fig:QFT-5} \Liquid circuit diagram for {\tt QFT'} on 5 qubits.}
\end{figure}


\textbf{Modular Addition. }
Modular exponentiation, that is the operation $a^r \mod(N)$ referred to above, can be performed using repeated multiplication, which in turn requires modular addition.
Here, we program a modular adder based on addition using the quantum Fourier transform \cite{Beauregard2002, Draper}.
In this design, both {\tt QFT} and {\tt QFT'} are required.
Throughout, the {\tt '} indicates inverse.
The circuit requires subcircuits (not shown here, see \cite{Beauregard2002}) for addition controlled by two qubits ({\tt CCAdd}), addition controlled by one qubit ({\tt CAddA}), and addition without controls ({\tt AddA}).
\definecolor{mygreen}{rgb}{0,0.3,0}
\definecolor{myred}{rgb}{0.5,0,0}
\lstset{language=Caml,
	  basicstyle=\scriptsize,
	  tabsize=4,
	  morecomment=[l]{//},
      keywordstyle=\color{myred},
      commentstyle=\color{mygreen},
	stringstyle=\color{mygreen},
	showspaces=false,
	showstringspaces=false,
	showtabs=false,
	morekeywords={CCAdd,AddA,QFT,CNOT,QFT,CAddA,CCAdd,X}}
\begin{lstlisting}
CCAdd a cbs 		// Perform the initial Add
AddA' N bs 			// Invert the add
QFT' bs				// Convert out of Fourier space
CNOT [bMx;anc] 		// Remember the overflow bit
QFT bs 				// Return to Fourier space
CAddA N (anc :: bs)	// Do the add based on overflow
CCAdd' a cbs 		// Undo the add
QFT' bs 			// Get out of Fourier space
X [bMx] 			// Use the top bit as a flag
CNOT [bMx;anc] 		// Clean up the Ancilla
X [bMx] 			// Revere use of the top bit
QFT bs 				// Return to Fourier space
CCAdd a cbs 		// Do the final version of the add
\end{lstlisting}

\begin{figure*}[t]
\centering
\input{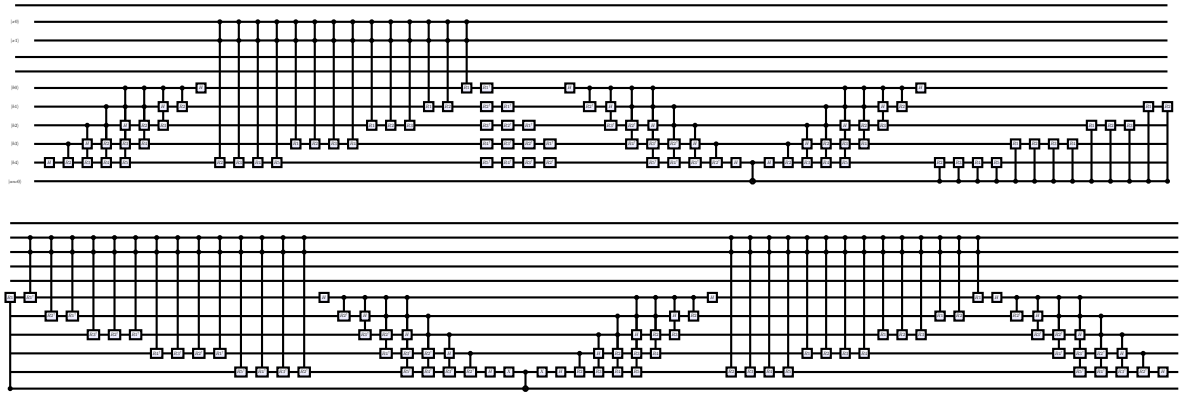}
\caption{\label{fig:AddModN-22} Part of the quantum circuit for modular addition generated by \LiquidB.}
\end{figure*}

Part of the circuit diagram for quantum modular addition is shown in Fig.~\ref{fig:AddModN-22}.

\subsection{Simulation}

To run the circuit for Shor's algorithm ({\tt Ua}) in \Liquid on inputs {\tt N} and {\tt a}, we can write:
\definecolor{mygreen}{rgb}{0,0.3,0}
\definecolor{myred}{rgb}{0.5,0,0}
\lstset{language=Caml,
	  basicstyle=\scriptsize,
	  tabsize=2,
	  morecomment=[l]{//},
      keywordstyle=\color{myred},
      commentstyle=\color{mygreen},
	stringstyle=\color{mygreen},
	showspaces=false,
	showstringspaces=false,
	showtabs=false,
	morekeywords={CompileUa,CacheStats,GrowPars,GrowGates,Dump,ShorRun,mapi,sum,Perms}}
\begin{lstlisting}
let circUa	= CompileUa N a qs	// Compile 1 Shor step
let count		= circUa.GateCount()*n*2
let hits,misses	=								// Get total gate count 
	Gate.CacheStats()							// Get gate caching stats
let gp			= GrowPars(30,2,false)		// Params for growing
let circUa	= circUa.GrowGates(k,gp)	// Grow the circuit
circUa.Dump()										// Dump circuit to file
ShorRun circUa rslt n a qs			// Run Shor
let m		= Array.mapi 						// Accumulate all the
	(fun i bit -> bit <<< i) rslt	// ..phase estimation bits
	|> Array.sum									// ..m = quantum result
let permG,permS,permN = k.Perms	// Get permutation stats
\end{lstlisting}

Here, {\tt N} is the number to be factored and the quantum circuit {\tt circUa} computes the order of {\tt a} modulo {\tt N}.
The input value {\tt a} is randomly chosen to be between {\tt 1} and {\tt N-1}.
The order {\tt rslt} is then used during classical post-processing (last 4 lines of the code) to either output a valid non-trivial factor of {\tt N} or to output failure.
Full statistics on the number of different quantum and classical gates may be obtained by running the command {\tt GateCount}.

Several rounds of Shor's algorithm may be required to find the factors of a number due to the algorithm's probabilistic nature.
As an example, say we want to factor the number 65. We can type:
\begin{alltt}\scriptsize
\emph{> Liquid __Shor(65,true)}
      65 = N = Number to factor
0.002676 = mins for compile
   43610 = cnt of gates
0.019366 = mins for growing gates
    1708 = cnt of gates
0.242003 = mins for running
   10675 = m = quantum result
 83.3984 = c =~ 10675/128
      64 = 128/2 = exponent
      62 = 32^64 + 1 mod 65
      60 = 32^64 - 1 mod 65
 GOT: 65 = 5x 13; n=7; mins=0.26; SUCCESS!!
\end{alltt}
In this circuit implementation (based on Beauregard's circuit \cite{Beauregard2002} for Shor’s algorithm), factoring 65 requires 17 qubits.
We compiled the circuit of 44,045 gates, compressed that down to 1,885 gates (by ``growing'' unitaries together), ran the result, and then performed the necessary classical post-processing.
All of this was done in a highly parallel fashion taking less than a minute.

The largest number we have factored is a 14-bit number ($8193$) which required 31 qubits in 50GB of memory, 28 rounds with half a million gates per round (reduced to 18,000 using gate growing), and ran for 43384 minutes ($30.1$ days). The answer was $8193 = 3 \times 2731$.  The simulation output is provided in Appendix \ref{app:Shor}.
This represents the largest number fully factored in a quantum computer simulator.\footnote{Authors of Ref.~\cite{Jugene} have shown the classical requirements for pieces of quantum factorization of a 15-bit number on a supercomputer.  We have fully factored a 14-bit number on a single desktop, simulating an end-to-end circuit implementation of Shor's algorithm.}
Factoring a 14-bit number is of course still within the range of instant solution in the classical realm; exponential scaling becomes important in the range at and beyond 1024-2048 bits, which represent current and future RSA key sizes.

These numbers are generated from a simulation (on a classical computer) of the quantum operations. 
A real quantum computer could factor this size instance in negligible time. 
The goal in \Liquid is to simulate all operations that would be performed on the quantum machine to enable algorithm development, optimization, and verification of correctness. Previous simulations have not factored numbers beyond 15 and 21, equivalent to 13 qubits and 70K gates (to the best of our knowledge). Our simulations, due to extensive optimization, can target simulations using up to around 30 qubits using only 32 GB RAM. The number 8193 required 31 qubits and 7M gates.

\begin{figure}[htb]
\centering
\includegraphics[width=3.5in]{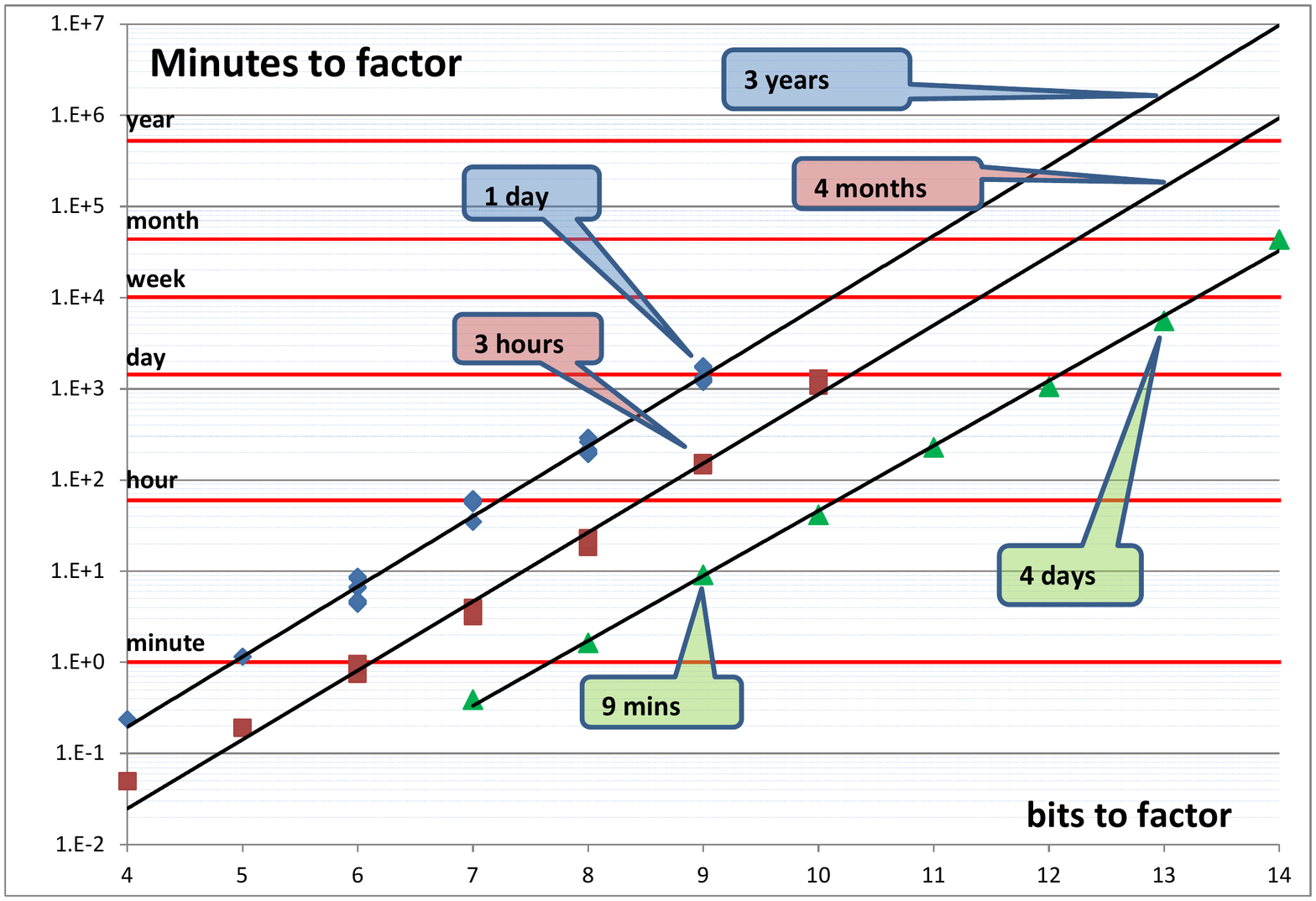}
\caption{\label{fig:ShorSim}Plot of number of bits vs.~simulation time (in minutes) for simulation of Shor's algorithm on varying number of qubits.}
\end{figure}

Fig.~\ref{fig:ShorSim} plots the \Liquid simulation time of Shor's algorithm for a range of bits.
The blue diamonds represent an early implementation with optimized linear algebra and simulation of each gate sequentially.
The red squares are after adding gate growing (massively reducing the number of gates).
The green triangles are after a full rewrite of the complex math package with optimized memory usage and tighter inner loops.
The significant improvements between the blue and green markers (from 3 years to 4 days for 13-bit simulation) highlights the importance of optimized simulation environments and domain-specific languages and tools for quantum computing.

\section{Conclusions and Future Work}\label{sec:Conclude}

\Liquid is a fully architected (Fig.~\ref{fig:arch}) quantum software platform that allows for efficient simulation of complex quantum circuits in a variety of different environments. It is designed as a modular system which makes it flexible and extensible by the user. A large number of gates are already provided, all of which may be overridden or extended. Three different classes of simulators are available as well as three different run times. Circuits may be defined and manipulated in many ways and may even be exported for running on various back-ends (both classical and quantum).

In future versions of \Liquid we plan to extend the software architecture to include layout of qubits, improved simulation of realistic noise models, gate timing constraints, and additional quantum error correction support.
We also plan to more closely integrate classical and quantum instructions and incorporate the ability to manipulate large sub-circuits, such as taking the adjoint of a circuit consisting of both classical and quantum pieces. A key extension will be the ability to specify architectural constraints such as timing, communication latency, and qubit proximity. Quantum algorithms with classical components may then be mapped to specific hardware implementations in the laboratory.  Soon, this will provide researchers with invaluable information for experimenting with future designs of quantum computers.


\begin{thebibliography}{28}
\providecommand{\natexlab}[1]{#1}
\providecommand{\url}[1]{\texttt{#1}}
\expandafter\ifx\csname urlstyle\endcsname\relax
  \providecommand{\doi}[1]{doi: #1}\else
  \providecommand{\doi}{doi: \begingroup \urlstyle{rm}\Url}\fi

\bibitem[Aaronson and Gottesman(2004)]{Aaronson2004}
S.~Aaronson and D.~Gottesman.
\newblock Improved simulation of stabilizer circuits.
\newblock \emph{Phys. Rev. A}, 70:\penalty0 052328, 2004.

\bibitem[Altenkirch and Green()]{Altenkirch2010}
T.~Altenkirch and A.~Green.
\newblock The quantum {IO} monad.
\newblock In S.~Gay and I.~McKie, editors, \emph{Semantic Techniques in Quantum
  Computation}, pages 173--205. Cambridge University Press.

\bibitem[Aspuru-Guzik et~al.(2005)Aspuru-Guzik, Dutoi, Love, and
  Head-Gordon]{AG2005}
A.~Aspuru-Guzik, A.~D. Dutoi, P.~J. Love, and M.~Head-Gordon.
\newblock Simulated quantum computation of molecular energies.
\newblock \emph{Science}, 309\penalty0 (5741):\penalty0 1704--1707, 2005.

\bibitem[Backus et~al.(1956)Backus, Beeber, Best, Goldberg, Herrick, Hughes,
  Mitchell, Nelson, Nutt, Sayre, Sheridan, Stern, and Ziller]{Fortran1957}
J.~Backus, R.~Beeber, S.~Best, R.~Goldberg, H.~Herrick, R.~Hughes, L.~Mitchell,
  R.~Nelson, R.~Nutt, D.~Sayre, P.~Sheridan, H.~Stern, and I.~Ziller.
\newblock The {FORTRAN} automatic coding system for the {IBM 704 EDPM}:
  Programmer's reference manual.
\newblock 1956.

\bibitem[Beauregard(2002)]{Beauregard2002}
S.~Beauregard.
\newblock Circuit for shor's algorithm using $2n+3$ qubits.
\newblock 2002.

\bibitem[Bettelli et~al.(2003)Bettelli, Calarco, and Serafini]{Bettelli2003}
S.~Bettelli, T.~Calarco, and L.~Serafini.
\newblock Toward an architecture for quantum programming.
\newblock \emph{European Physics D}, 25\penalty0 (2):\penalty0 181--200, 2003.

\bibitem[Deutsch(1985)]{Deutsch1985}
D.~Deutsch.
\newblock Quantum theory, the {C}hurch-{T}uring principle, and the universal
  quantum computer.
\newblock \emph{Proc. R. Soc. Lond. A}, 400\penalty0 (97), 1985.

\bibitem[Draper(2000)]{Draper}
T.~G. Draper.
\newblock Addition on a quantum computer.
\newblock 2000.

\bibitem[Farhi et~al.(2000)Farhi, Goldstone, Gutmann, and Sipser]{Farhi}
E.~Farhi, J.~Goldstone, S.~Gutmann, and M.~Sipser.
\newblock Quantum computation by adiabatic evolution.
\newblock 2000.

\bibitem[Garcia and Markov(2013)]{Garcia2013}
H.~J. Garcia and I.~L. Markov.
\newblock Quipu: High-performance simulation of quantum circuits using
  stabilizer frames.
\newblock In \emph{Intl. Conf. Computer Design, ICCD}, pages 404--410, 2013.

\bibitem[Green et~al.(2013)Green, Lumsdaine, Ross, Selinger, and
  Valiron]{Selinger2013}
A.~Green, P.~L. Lumsdaine, N.~Ross, P.~Selinger, and B.~Valiron.
\newblock Quipper: A scalable quantum programming language.
\newblock In \emph{Proceedings of PLDI '13}, 2013.

\bibitem[Harrow et~al.(2009)Harrow, Hassidim, and Lloyd]{HHL2009}
A.~W. Harrow, A.~Hassidim, and S.~Lloyd.
\newblock Quantum algorithm for solving linear systems of equations.
\newblock \emph{Phys. Rev. Lett.}, 15\penalty0 (3):\penalty0 150502, 2009.

\bibitem[Jordan()]{JordanZoo}
S.~Jordan.
\newblock Quantum algorithm zoo.
\newblock http://math.nist.gov/quantum/zoo/.

\bibitem[Knill(1996)]{Knill1996}
E.~Knill.
\newblock Conventions for quantum pseudocode.
\newblock Technical Report LAUR-96-2724, LANL, 1996.

\bibitem[Lanyon et~al.(2009)Lanyon, Whitfield, Gillet, Goggin, Almeida, Kassal,
  Biamonte, Mohseni, Powell, Barbieri, Aspuru-Guzik, and White]{Chem2009}
B.~P. Lanyon, J.~D. Whitfield, G.~G. Gillet, M.~E. Goggin, M.~P. Almeida,
  I.~Kassal, J.~D. Biamonte, M.~Mohseni, B.~J. Powell, M.~Barbieri,
  A.~Aspuru-Guzik, and A.~G. White.
\newblock Towards quantum chemistry on a quantum computer.
\newblock \emph{Nature Chemistry}, 2:\penalty0 106--111, 2009.

\bibitem[Miszczak(2011)]{Miszczak2011}
J.~Miszczak.
\newblock Models of quantum computation and quantum programming languages.
\newblock \emph{Bull. Pol. Acad. Sci.-Tech. Sci.}, 59\penalty0 (3):\penalty0
  305--324, 2011.

\bibitem[Nielsen and Chuang(2000)]{Nielsen2000}
M.~A. Nielsen and I.~L. Chuang.
\newblock \emph{{Quantum Computation and Quantum Information}}.
\newblock Cambridge University Press, 2000.
\newblock ISBN 0521635039.

\bibitem[\"{O}mer(1998)]{Omer1998}
B.~\"{O}mer.
\newblock A procedural formalism for quantum computing.
\newblock Master's thesis, Theoretical University of Vienna, 1998.

\bibitem[\"{O}mer(2000)]{Omer2000}
B.~\"{O}mer.
\newblock Quantum programming in {QCL}.
\newblock Master's thesis, Technical University of Vienna, 2000.

\bibitem[\"{O}mer(2003)]{Omer2003}
B.~\"{O}mer.
\newblock \emph{Structured Quantum Programming}.
\newblock PhD thesis, Theoretical University of Vienna, 2003.

\bibitem[Raedt et~al.(2007)Raedt, Michielsenb, Raedt, Trieuc, Arnold, Richter,
  Lippert, Watanabe, and Ito]{Jugene}
K.~D. Raedt, K.~Michielsenb, H.~D. Raedt, B.~Trieuc, G.~Arnold, M.~Richter,
  T.~Lippert, H.~Watanabe, and N.~Ito.
\newblock Massively parallel quantum computer simulator.
\newblock \emph{Comp. Phys. Comm.}, 176:\penalty0 121--136, 2007.

\bibitem[Selinger and Valiron(2006)]{Selinger2006}
P.~Selinger and B.~Valiron.
\newblock A lambda calculus for quantum computation with classical control.
\newblock \emph{Mathematical Structures in Computer Science}, 16\penalty0
  (3):\penalty0 527--552, 2006.

\bibitem[Selinger and Valiron(2009)]{Selinger2009}
P.~Selinger and B.~Valiron.
\newblock Quantum lambda calculus.
\newblock pages 135--172. Cambridge University Press, 2009.

\bibitem[Shor(1997)]{Shor1994}
P.~Shor.
\newblock Polynomial-time algorithms for prime factorization and discrete
  logarithms on a quantum computer.
\newblock \emph{SIAM Journal of Computing}, 26:\penalty0 1484--1509, 1997.

\bibitem[Svore et~al.(2006)Svore, Aho, Cross, Chuang, and Markov]{Svore2006}
K.~M. Svore, A.~V. Aho, A.~Cross, I.~Chuang, and I.~Markov.
\newblock A layered software architecture for quantum computing design tools.
\newblock \emph{Computer}, 39\penalty0 (1):\penalty0 74--83, 2006.

\bibitem[{van Tonder}(2004)]{Tonder2004}
A.~{van Tonder}.
\newblock A lambda calculus for quantum computation.
\newblock \emph{SIAM Journal of Computation}, 33\penalty0 (5):\penalty0
  1109--1135, 2004.

\bibitem[Viamontes et~al.(2005)Viamontes, Markov, and Hayes]{Viamontes2005}
G.~Viamontes, I.~Markov, and J.~Hayes.
\newblock Graph-based simulation of quantum computation in the state-vector and
  density-matrix representation.
\newblock \emph{Quantum Information and Computation}, 5\penalty0 (2):\penalty0
  113--130, 2005.

\bibitem[Yao(1993)]{Yao1993}
A.~Yao.
\newblock Quantum circuit complexity.
\newblock In \emph{Proc. of the 34th IEEE Symposium on Foundations of Computer
  Science}, pages 352--360. IEEE Press, 1993.

\end{thebibliography}

\appendix
\section*{APPENDIX}
\setcounter{section}{0}

\section{Quantum Gates}\label{app:Gates}
\begin{figure*}[ht!]
\centering
\includegraphics[width=7in]{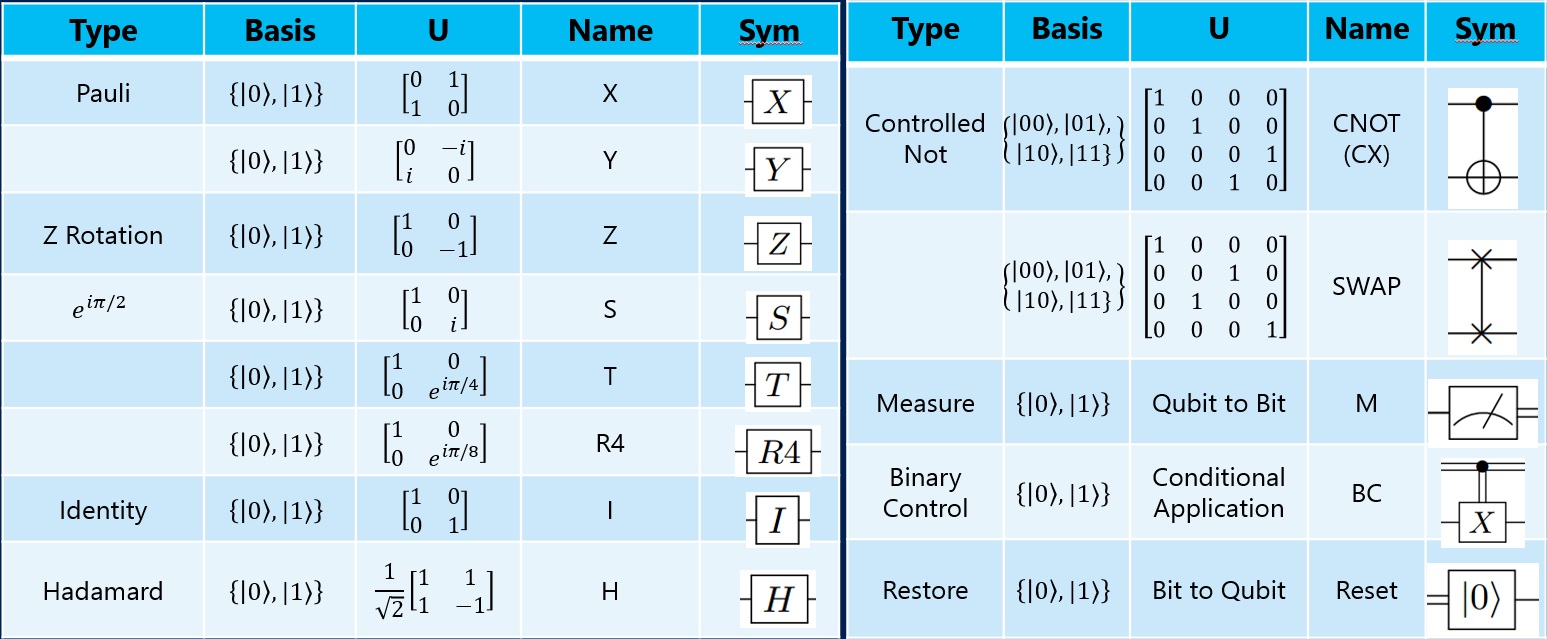}
\caption{\label{app:gates}Sampling of basic quantum gates available in \Liquid.}
\end{figure*}

Standard gates are presented in Figure \ref{app:gates}.

\section{Large Shor Simulation Run}\label{app:Shor}
This is the raw log from factoring the number $8193$ (14 bits, 31 qubits) in 30.1 days. This includes 28 applications of a Shor round (each defined with 515,032 gates). The section starting with ``Wrap circuit pieces'' is the gate growing (reducing down to 18,200 gates to simulate, some of which were matrices that spanned 30 qubits ($2^{30}\times2^{30}$). Each of the 28 rounds is shown with the single bit result ({\tt m=}) for that round. At the end is the classical post-processing that generates the factors ({\tt 3 x 2731}).

\begin{tiny}
\begin{verbatim}
0:0000.0/=============== Logging to: Liquid.log opened ================
0:0000.0/======== Doing Shor Round =========
0:0000.0/    8193 = N = Number to factor
0:0000.0/    1024 = a = coPrime of N
0:0000.0/      14 = n = number of bits for N
0:0000.0/   16384 = 2^n
0:0000.0/      31 = total qubits
0:0000.0/      23 = starting memory (MB)
0:0000.0/         - Compiling circuit
0:0000.0/Qubits: M at  0
0:0000.0/Qubits: X from  1 to 13
0:0000.0/Qubits: B from 14 to 28
0:0000.0/Qubits: Anc at 30
0:0000.0/0.004982 = mins for compile
0:0000.0/  515032 = cnt of gates
0:0000.0/   58253 = cache hits
0:0000.0/     275 = cache misses
0:0000.0/      31 = compiled memory (MB)
0:0000.0/         - Wrapping circuit pieces
0:0000.0/            8 wires, possibles: 2935 (did=     0 big=     0)
0:0000.0/            9 wires, possibles: 2527 (did=   408 big=   116)
0:0000.1/           10 wires, possibles: 1870 (did=  1065 big=   232)
0:0000.1/           11 wires, possibles: 1534 (did=  1401 big=   348)
0:0000.1/           12 wires, possibles: 1310 (did=  1625 big=   494)
0:0000.1/           13 wires, possibles: 1216 (did=  1719 big=   726)
0:0000.1/           14 wires, possibles:  979 (did=  1956 big=   958)
0:0000.2/           15 wires, possibles:  732 (did=  2203 big=  1192)
0:0000.2/           16 wires, possibles:  696 (did=  2239 big=  1568)
0:0000.3/           17 wires, possibles:  677 (did=  2258 big=  2069)
0:0000.3/           18 wires, possibles:  650 (did=  2285 big=  2573)
0:0000.3/           19 wires, possibles:  650 (did=  2285 big=  3178)
0:0000.3/           20 wires, possibles:  650 (did=  2285 big=  3805)
0:0000.3/           21 wires, possibles:  650 (did=  2285 big=  4441)
0:0000.3/           22 wires, possibles:  650 (did=  2285 big=  5080)
0:0000.3/           23 wires, possibles:  650 (did=  2285 big=  5721)
0:0000.3/           24 wires, possibles:  650 (did=  2285 big=  6363)
0:0000.3/           25 wires, possibles:  650 (did=  2285 big=  7006)
0:0000.3/           26 wires, possibles:  650 (did=  2285 big=  7650)
0:0000.3/           27 wires, possibles:  650 (did=  2285 big=  8294)
0:0000.3/           28 wires, possibles:  650 (did=  2285 big=  8938)
0:0000.3/           29 wires, possibles:  650 (did=  2285 big=  9584)
0:0000.3/           30 wires, possibles:  650 (did=  2285 big= 10230)
0:0000.3/           31 = Ran out of wires
0:0000.3/           MM: g:  2285 b: 10876  17=27 16=19 15=36 14=247 13=237
0:0000.3/0.291275 = mins for growing gates
0:0000.3/   18200 = cnt of gates
0:0000.3/    1184 = grown memory (MB)
0:0000.3/         - Running circuit
0:0000.3/Qubits: M at  0
0:0000.3/Qubits: X from  1 to 13
0:0000.3/Qubits: B from 14 to 28
0:0000.3/Qubits: Anc at 30
0:0912.5/          1 of 28 [MB:17857 m=1]
0:2520.8/          2 of 28 [MB:17964 m=1]
0:4085.6/          3 of 28 [MB:18034 m=1]
0:5647.8/          4 of 28 [MB:18115 m=1]
0:7221.3/          5 of 28 [MB:18195 m=0]
0:8770.2/          6 of 28 [MB:18276 m=0]
0:10329.0/          7 of 28 [MB:18356 m=0]
0:11872.3/          8 of 28 [MB:18436 m=0]
0:13426.4/          9 of 28 [MB:18517 m=1]
0:14981.8/         10 of 28 [MB:18614 m=0]
0:16576.0/         11 of 28 [MB:18710 m=0]
0:18162.4/         12 of 28 [MB:18806 m=0]
0:19703.0/         13 of 28 [MB:18903 m=1]
0:21310.8/         14 of 28 [MB:19007 m=1]
0:22889.2/         15 of 28 [MB:19097 m=1]
0:24468.6/         16 of 28 [MB:19199 m=0]
0:26071.0/         17 of 28 [MB:19289 m=1]
0:27648.6/         18 of 28 [MB:19386 m=0]
0:29218.8/         19 of 28 [MB:19490 m=1]
0:30759.1/         20 of 28 [MB:19595 m=0]
0:32339.7/         21 of 28 [MB:19700 m=1]
0:33891.9/         22 of 28 [MB:19804 m=1]
0:35479.4/         23 of 28 [MB:19908 m=0]
0:37069.8/         24 of 28 [MB:20019 m=0]
0:38627.6/         25 of 28 [MB:20117 m=0]
0:40180.1/         26 of 28 [MB:20222 m=1]
0:41767.6/         27 of 28 [MB:20326 m=1]
0:43384.4/         28 of 28 [MB:20430 m=1]
0:43384.4/43383.055681 = mins for running
0:43384.4/2.603e+06 = Elapsed time (seconds)
0:43384.4/      31 = Max Entangled
0:43384.4/       0 = Gates Permuted
0:43384.4/   18199 = State Permuted
0:43384.4/     115 = None  Permuted
0:43384.4/238383375 = m = quantum result
0:43384.4/ 14549.8 = c =~ 238383375/16384 
0:43384.4/    8192 = 16384/2 = exponent
0:43384.4/    8066 = 1024^8192 + 1 mod 8193
0:43384.4/    8064 = 1024^8192 - 1 mod 8193
0:43384.4/GOT: 8193=   3x2731 co= 1024 n,q=14,31 mins=43383.35 SUCCESS!!
0:43384.4/=============== Logging to: Liquid.log closed ================
\end{verbatim}
\end{tiny}

\section{\Liquid Built-in Tests}\label{app:Tests}

\begin{tiny}
\begin{verbatim}
Big()             Try to run large entanglement tests (20 through 33 qubits)
Chem(n,t,b,o,c)   Test n (try 99 for help) and then H2O params
Correct()         Use 15 qubits and random circuits to test teleport in several ways
EIGS()            Check eigevalues using ARPACK
Entangle1()       Draw and run 24 qubit entanglement circuit
Entangles()       Draw and run 100 instances of 16 qubit entanglement test
EntEnt()          Entanglement entropy test
EPR()             Draw EPR circuit (.svg files)
Ferro(false,true) Test ferro magnetic coupling with true=full, true=runonce
H2()              Solve ground state for H2 molecule
H2O(t,b,o,c)      Solve ground state for H2O (trotter=32,bits=20,order=1,2,coal=-1.0 or <=1.0)
Hubbard("pars")   Hubbard model (basic test, use "INIT JOIN" for pars), see docs for more
MPS(bMn,bInc,bMx) Run an MPS simulation of a ferro chain typically between B=0.0 and 2.0
MPS1(h,B,bd,acc)  Run an MPS simulation of a ferro chain (h=0,B=1.0 is the critical point)
Noise1(d,i,p)     Noise on 1 qubit. depth,iters,probOfNoise
NoiseAmp()        Amplitude damping (non-unitary) noise
NoiseTele(S,i,p)  Noise on Teleport S=doSteane? i=iters p=prob
QECC()            Test teleport with error injection in Steane7 code (output drawings)
QFTbench()        Bench mark various execution modes for QFT (in Shor package)
QuAM()            Quantum Associative Memory
QWalk(typ)        Walk tiny,tree,graph or RMat file with graph information
Ramsey33()        Try to find a Ramsey(3,3) solution
SG()              Test spin glass model
Shor(15,true)     Factor N using Shor's algorithm false=direct true=optimized circuit
ShorT(true)       Draw and test each of the sub-operations in Shor false=direct true=circuit
show(str)         Test routine to echo str and then exit
Steane7()         Test basic error injection in Steane7 code
Teleport()        Draw and run original, circuit and grown versions
TSP(5)            Try to find a Traveling Salesman soltion for 5 to 8 cities
Vbasis(eps)       Test Vbasis generation with eps (1.e-20 typical)
\end{verbatim}
\end{tiny}

\section{Hamiltonian Simulation}\label{app:Ham}

A package for simulating Hamiltonians is included in \Liquid and built on top of the universal modeling simulator.
There are three main ways to use this environment.

\subsection{Adiabatic simulator}\label{sec:Spin}
The first is with time-varying Hamiltonians that represent adiabatic spin glass problems (\ref{eqn:spin}). This simulator has been used for applications from modeling the D-Wave machine (a hardware decoherence model is available) to implementing Machine Learning algorithms (e.g., Traveling Salesman).

\begin{equation}
H=\Gamma(t)\sum_i \Delta_i \sigma^x_i + \Lambda(t) \left( \sum_i h_i\sigma^z_i + \sum_{i<j} J_{ij} \sigma^z_i \sigma^z_j \right)
\label{eqn:spin}
\end{equation}

The adiabatic approach starts in a known ground state in $\sigma_x$ and then moves continuously to the unknown ground state in $\sigma_z$ (which is the solution of our problem). By moving slowly enough we can stay in the ground state of the entire system and reach the solution to the problem specified by the $h_i$ and $J_{ij}$ values in the equation.

\subsection{Fermionic simulator}\label{sec:Ferm}
\begin{figure}[h]
\centering
\includegraphics{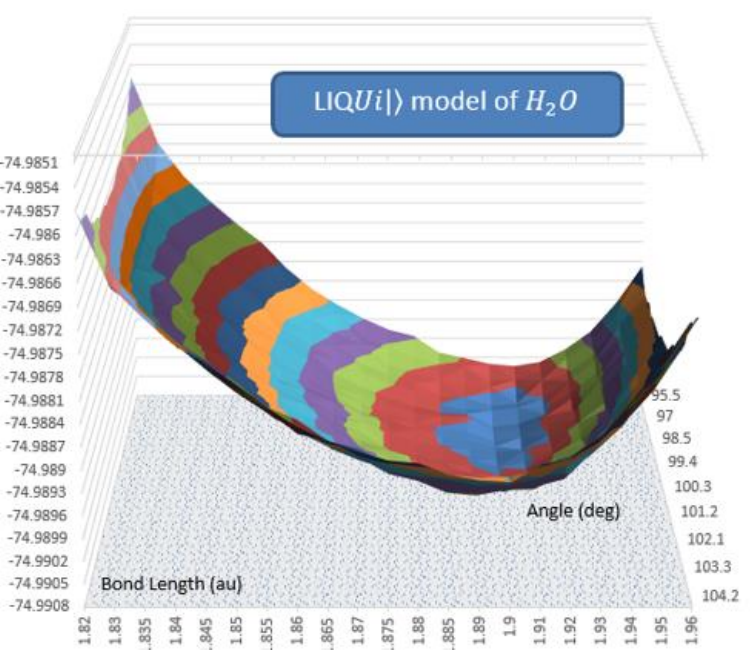}
\caption{\label{fig:h2o}Results of $H_2O$ ground state modeling}
\end{figure}

The fermionic Hamiltonian (\ref{eqn:ferm}) is a second quantized Hamiltonian that represents the interactions of electrons in a molecular model. \Liquid provides gates that represent number, excitation, Coulomb, exchange, number excitation and double excitation operators. This simulator has been used to implement sophisticated models including ones for H$_2$ and H$_2$O. Fig.~\ref{fig:h2o} shows a complete ground state model for water where the x axis varies the bond length between the oxygen and the hydrogen atoms, while the y axis varies the angle between the hydrogen bonds. The z axis is the energy predicted (units are Hartree).

\begin{equation}
H=\sum_{p<q}h_{pq}a^\dag_p a_q + \frac{1}{2}\sum_{p<q<r<s}h_{pqrs}a^\dag_p a^\dag_q a_r a_s
\label{eqn:ferm}
\end{equation}

The first half of the equation represent the single electron terms ({\tt Hpp Hpq}) while the second half are the two electron terms ({\tt Hpqqp Hpqqr Hpqrs}).

\subsection{Mixing simulators}\label{sec:Mixed}
The adiabatic and fermionic simulators can be mixed to allow fermionic simulation of time-varying Hamiltonians. One example of this is implemented as the Hubbard model (\ref{eqn:hub}) which is an effective Hamiltonian for modeling high temperature superconductors (cuprates).

\begin{eqnarray}
H = & -\sum_{<i,j>}\sum_\sigma t_{ij} \left( c^\dag_{i,\sigma}c_{j,\sigma} +
        c^\dag_{j,\sigma}c_{i,\sigma} \right) + \nonumber\\
    & U\sum_i \eta_{i,\uparrow}\eta_{i,\downarrow} + \sum_i \epsilon_i \eta_i
    \label{eqn:hub} \\
    \nonumber \\
    & \eta_{i,\sigma} = c^\dag_{i,\sigma}c_{i,\sigma} = \textit{local spin density} \nonumber \\
    & \eta_i = \sum_\sigma \eta_{i,\sigma} = \textit{total local density} \nonumber
\end{eqnarray}

\begin{figure}[h]
\centering
\includegraphics[scale=0.6]{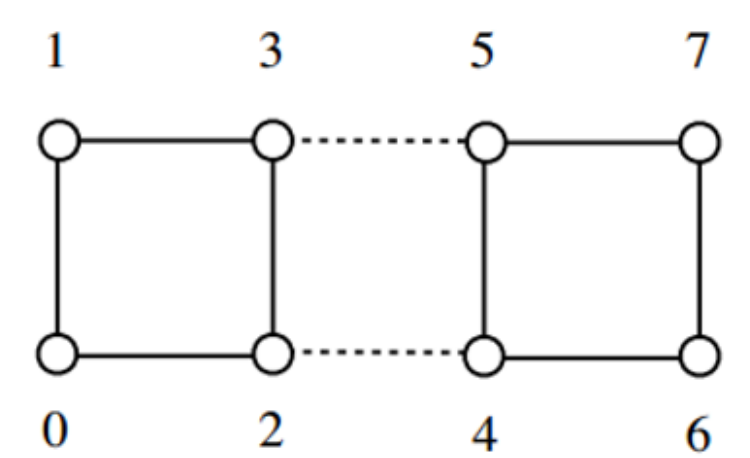}
\caption{\label{fig:lat}Hubbard lattice model with two plaquettes}
\end{figure}

The model implemented is a 2d lattice (as shown in Fig.~\ref{fig:lat}) where we define plaquettes that will be evolved adiabatically separately into the ground state, merged, and then separated to determine if we are left in a superconducting state.

\begin{figure}[h]
\includegraphics[scale=0.8]{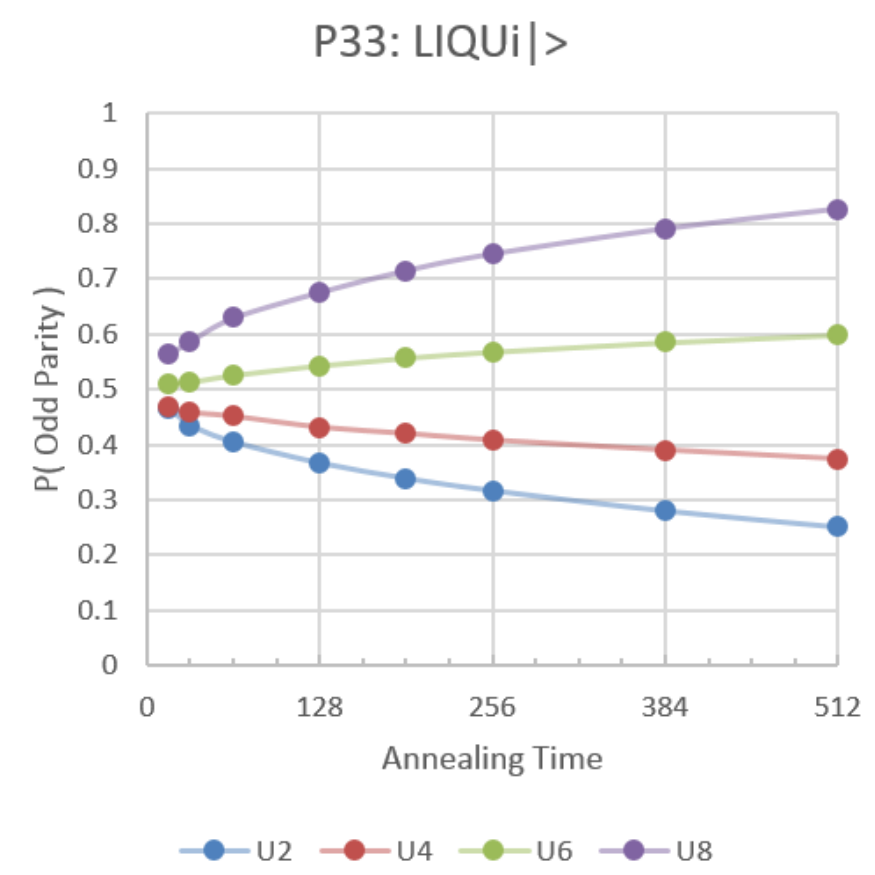}
\caption{\label{fig:p33}Probability of breaking superconducting pairs as function of annealing time for various interaction strengths.}
\end{figure}

The goal is to model different chemical compositions as spacing between the copper oxide layers in a cuprate, which modify the ratio of interaction to hopping $U/t$. After preparing 2 plaquettes with 6 electrons we adiabatically separate them and measure the probability of finding three electrons on each (Fig.~\ref{fig:p33}). If electrons are paired, the probability of having an odd number should be suppressed, and we can thus see pairing as suppression of $P_{33}$ in the figure as the length of our annealing schedule is increased.






\end{document}